\documentstyle[12pt,aaspp4,flushrt]{article}

\singlespace

\font\twelvemib=cmmib10 scaled 1200 \font\tenmib=cmmib10
\font\twelvembsy=cmbsy10 scaled 1200 \font\tenmbsy=cmbsy10

\newfam\mibfam
\textfont\mibfam=\twelvemib \scriptfont\mibfam=\tenmib
\scriptscriptfont\mibfam=\tenmib
\def\mbf{\fam\mibfam}
\newfam\mbsyfam
\textfont\mbsyfam=\twelvembsy \scriptfont\mbsyfam=\tenmbsy
\scriptscriptfont\mbsyfam=\tenmbsy

\mathchardef\alpha="710B
\mathchardef\beta="710C
\mathchardef\gamma="710D
\mathchardef\delta="710E
\mathchardef\epsilon="710F
\mathchardef\zeta="7110
\mathchardef\eta="7111
\mathchardef\theta="7112
\mathchardef\iota="7113
\mathchardef\kappa="7114
\mathchardef\lambda="7115
\mathchardef\mu="7116
\mathchardef\nu="7117
\mathchardef\xi="7118
\mathchardef\pi="7119
\mathchardef\rho="711A
\mathchardef\sigma="711B
\mathchardef\tau="711C
\mathchardef\upsilon="711D
\mathchardef\phi="711E
\mathchardef\chi="711F
\mathchardef\psi="7120
\mathchardef\omega="7121
\mathchardef\varepsilon="7122
\mathchardef\vartheta="7123
\mathchardef\varpi="7124
\mathchardef\varrho="7125
\mathchardef\varsigma="7126
\mathchardef\varphi="7127
\mathchardef\nabla="7272
\mathchardef\cdot="7201

\def\vct#1{{\mbf #1}}

\def\matrixmark#1{\widehat{\cal #1}}
\def\shear{{\matrixmark S}}
\def\tshear{{\matrixmark T}}
\def\etal{{et~al.}}
\def\refrule{\makebox[3pc]{\leaders\hrule depth-2pt height 2.4pt\hfill}}

\lefthead{Lee et al.}
\righthead{Statistics of Microlensing Magnification. II.}

\begin{document}

\hyphenation{macro-im-age mi-cro-im-age}

\title{STATISTICS OF GRAVITATIONAL MICROLENSING MAGNIFICATION.\
       II.\ THREE-DIMENSIONAL\\ LENS DISTRIBUTION}

\author{Man Hoi Lee}
\affil{Department of Physics, Queen's University, Kingston,
       Ontario K7L 3N6, Canada}
\authoremail{mhlee@astro.queensu.ca}

\author{Arif Babul}
\affil{Department of Physics, New York University,
       4 Washington Place, New York, NY 10003}
\authoremail{babul@almuhit.physics.nyu.edu}

\author{Lev Kofman}
\affil{Institute for Astronomy, University of Hawaii,
       2680 Woodlawn Drive, Honolulu, HI 96822}
\authoremail{kofman@ifa.hawaii.edu}

\and

\author{Nick Kaiser}
\affil{CIAR Cosmology Program,
       Canadian Institute for Theoretical Astrophysics,
       University of Toronto, 60 St.~George Street, Toronto,
       Ontario M5S 1A7, Canada}
\authoremail{kaiser@cita.utoronto.ca}

\bigskip

\begin{abstract}
In Paper I we studied the theory of gravitational microlensing for a planar
distribution of point masses.
In this second paper,
we extend the analysis to a three-dimensional lens distribution.
First we study the lensing properties of three-dimensional lens
distributions by considering in detail the critical curves, the caustics, the
illumination patterns, and the magnification cross-sections $\sigma(A)$ of
multiplane configurations with two, three, and four point masses.
For $N_\ast$ point masses that are widely separated in Lagrangian space,
we find that there are $\sim 2^{N_\ast} - 1$ critical curves in total,
but that only $\sim N_\ast$ of these produce prominent caustic-induced
features at the high magnification end of $\sigma(A)$.

In the case of a low optical depth random distribution of point masses,
we show that the multiplane lens equation near a point mass can be reduced
to the single plane equation of a point mass perturbed by weak shear.
This allows us to calculate the caustic-induced feature in the macroimage
magnification distribution $P(A)$ as a weighted sum of the semi-analytic
feature derived in Paper I for a planar lens distribution.
The resulting semi-analytic caustic-induced feature is similar to the
feature in the planar case, but it does not have any simple scaling
properties, and it is shifted to higher magnification.
The semi-analytic distribution is compared to the results of previous
numerical simulations for optical depth $\tau \sim 0.1$, and they are in
better agreement than a similar comparison in the planar case.
We explain this by estimating the fraction of caustics of individual lenses
that merge with those of their neighbors.
For $\tau = 0.1$, the fraction is $\approx 20\%$, much less than the
$\approx 55\%$ for the planar case.
\end{abstract}


\section{INTRODUCTION}

The gravitational lensing effect of an ensemble of point masses has long
been a subject of investigation (see, e.g., Schneider, Ehlers, \& Falco 1992
for a summary of earlier work).
Most of the studies have focused on the situation where the spatial
distribution of the point mass lenses is ``compact,'' i.e., the size of the
region within which the lenses are distributed is much smaller than the
other distances (those between the source and the lenses and between the
observer and the lenses) in the problem.
For gravitational lensing calculations, a compact distribution of lenses is
usually approximated by its projection onto a single lens plane.
The important case of low optical depth, single plane microlensing has been
investigated in great detail (Kaiser 1992; Mao 1992; Rauch \etal\ 1992;
Kofman \etal\ 1996, hereafter Paper I; see also Schneider \etal\ 1992 and
references therein),
and we now understand reasonably well the caustic structure, the
illumination pattern, the probability distribution of macroimage
magnification, and the flux variability of macroimages in this case.

There are, however, situations in which the lensing objects are modeled by
an ``extended'' (or three-dimensional) distribution of point masses, i.e.,
by a point mass distribution that is comparable in scale to the other
distances in the problem.
A well-known example is the microlensing of stars in the Large Magellanic
Cloud and the Galactic bulge by foreground stars and possibly massive
compact halo objects (MACHOs) (Alcock \etal\ 1993; Aubourg \etal\ 1993;
Udalski \etal\ 1993).
The theoretical analysis of this problem is relatively simple because the
optical depth to microlensing is very small ($\tau < 10^{-5}$) and the
point mass lenses are essentially acting independently
(Paczy\'nski 1986, 1991; Griest 1991).
(Note, however, that microlensing by physical binary systems is also
possible; Mao \& Paczy\'nski 1991; Udalski \etal\ 1994; Alard, Mao, \&
Guibert 1995.)
A more complicated problem is the possibility that a substantial fraction
of the dark matter in the universe is in the form of compact objects
(e.g., Carr 1994).
Then the universe as a whole has a significant optical depth to microlensing
(Press \& Gunn 1973; Dalcanton \etal\ 1994).
This latter problem has been studied by several authors using numerical
simulations (Refsdal 1970; Schneider \& Weiss 1988a,b; Rauch 1991), and
it was found that there are qualitative differences between two- and
three-dimensional microlensing.
For example, for $\tau \sim 0.1$, the three-dimensional macroimage
magnification distribution does not show the prominent caustic-induced
feature produced by a planar distribution of point masses
(Rauch \etal\ 1992; but see \S~4 below).

The usual approach to gravitational lensing by a three-dimensional
distribution of lenses is the multiple lens-plane approximation
(Blandford \& Narayan 1986; Kovner 1987).
The properties of multiplane gravitational lensing and, in particular, the
reasons for many of the differences between two- and three-dimensional
microlensing have not been investigated in detail, and they are not yet
well understood.
Besides the statistical microlensing studies mentioned above,
there have been few studies of the basic mathematical properties of
multiplane gravitational lensing (Levine \& Petters 1993; Levine, Petters,
\& Wambsganss 1993) or the properties of specific multiplane configurations
(Kochanek \& Apostolakis 1988; Lee \& Paczy\'nski 1990; Erdl \& Schneider
1993, hereafter ES).
This was perhaps due to the lack of lensing candidates with lenses at
different redshifts.
However, recent observations of the lens systems 2345$+$007 (Pell\'o \etal\
1996), 2016$+$112 (Lawrence, Neugebauer, \& Matthews 1993), and 0957$+$561
(Bernstein, Tyson, \& Kochanek 1993; but see Dahle, Maddox, \& Lilje 1994)
suggest that these systems may involve lenses at different redshifts.

As we shall see in \S~3, to understand some of the differences between two-
and three-dimensional microlensing, it is useful to study in detail
specific cases of simple three-di\-men\-sional lens configurations.
The study by Erdl \& Schneider (ES) of the lens system consisting of two
point masses at different distances from the observer is a step in this
direction.
They considered primarily the critical curves and the caustics and were
able to obtain a complete classification of their topological properties
with respect to the model parameters in this particular case.

The main motivation of this study is to give a systematic theory of
gravitational microlensing by a random distribution of point masses.
In the accompanying Paper I we refined the theory of gravitational
microlensing for a planar distribution of point masses.
In particular, we obtained a practical semi-analytic expression for the
macroimage magnification distribution, $P(A)$, for low optical depth $\tau$,
by allowing for the shear perturbations of neighboring lenses.
In this paper we study the properties of gravitational microlensing by a
three-dimensional distribution of point masses.
We are interested in the critical curves, the caustics, the illumination
pattern, and especially the macroimage magnification distribution.
We shall not consider other interesting aspects such as time delays and the
temporal behavior of microlensing events
(note, however, that the latter is given by the magnification along tracks
through the illumination pattern).
In \S~2 we write down the lens equations.
In \S~3 we study in detail multiplane configurations with two,
three, and four point masses to find out what new features are expected
in three-dimensional microlensing.
We also explain qualitatively why these new features arise.
The case of low optical depth three-dimensional microlensing is studied
in \S~4, using an approach similar to that used in Paper I to study the
caustic-induced feature in $P(A)$ for the planar case.
The resulting semi-analytic $P(A)$ is compared to the numerical results
of Rauch (1991), and the differences between the two- and three-dimensional
results are discussed.
Our conclusions are summarized in \S~5.

\section{LENS EQUATIONS}

We begin by establishing the relevant equations for gravitational lensing by
a three-dimensional lens distribution.
As in Paper I, we shall only consider gravitational lensing in an
Einstein-de Sitter cosmological background weakly perturbed by the
gravitational field of the lenses.
Our notation is identical to that used in Paper I, and the readers are
referred to \S~2 of Paper I for the definitions of terms not explicitly
defined here.

Let us consider an isotropic point source at the origin of our coordinates
and an observer plane at a comoving distance $\chi_{so}$ from the source
(see Fig.~1).
In the absence of perturbations, the observer plane would be
uniformly illuminated and a light ray which leaves the source with angle
$\vct\theta = (\theta_1, \theta_2)$ would pierce the observer plane
at $\vct x = \chi_{so} \vct \theta$.
This defines our planar Lagrangian coordinates $\vct x$.
In the presence of density inhomogeneities, the light ray would suffer
deflections and pierce the observer plane at the Eulerian coordinates
$\vct r$.

For a three-dimensional (or extended) lens distribution, gravitational
lensing is usually studied within the framework of the multiple lens-plane
approximation (Blandford \& Narayan 1986; Kovner 1987).
It is assumed that the lens distribution can be approximated by a series
of $N$ geometrically thin configurations, i.e., lens planes,
and that the separations between the lens planes are sufficiently large.
Then the deflection of a light ray by one of the lens planes is
not influenced by the others and a light ray is deflected $N$ times on the way
from the source to the observer (see Fig.~1).
After $j-1$ deflections, the location at which the light ray pierces the
$j$th lens plane, projected onto the observer plane, is
\begin{equation}
{\vct x}_j = {\vct x}_1 + \sum_{i=1}^{j-1} \beta_{ij} {\vct s}_i({\vct x}_i),
\label{eq1}
\end{equation}
where ${\vct s}_i ({\vct x}_i)$ is the deflection by the $i$th plane
(which can be expressed as the gradient with respect to ${\vct x}_i$ of an
effective surface gravitational potential on plane $i$),
$\beta_{ij}$ is a dimensionless distance parameter defined by
\begin{equation}
\beta_{ij} \equiv {\chi_{ij}\chi_{so} \over \chi_{sj}\chi_{io}},
\label{eq2}
\end{equation}
and $\chi_{ij}$ is the comoving distance between planes $i$ and $j$.
Note that $\beta_{ij}$ can take on values between 0 (in the limit $i=j$) and
1 (in the limit $i=s$ or $j=o$).
If we denote the observer plane as plane $N+1$, then the Lagrangian to
Eulerian mapping is
\begin{mathletters}
\begin{equation}
\vct r = \vct x + \vct s ,
\end{equation}
where
\begin{equation}
\vct r = {\vct x}_{N+1}, \qquad \vct x = {\vct x}_1, \qquad
\vct s = \sum_{i=1}^{N} {\vct s}_i ({\vct x}_i).
\end{equation}
\end{mathletters}
Unlike the single plane lens mapping, this multiplane lens mapping is not
in general a gradient mapping.

For the particular case where the lens planes are populated by point masses,
\begin{equation}
{\vct s}_i ({\vct x}_i) = -4G \left({\chi_{io} \chi_{so} \over
                                     a_i \chi_{si}}\right)
  \sum_k m_{ik} {{\vct x}_i-{\vct x}_{ik} \over |{\vct x}_i-{\vct x}_{ik}|^2},
\end{equation}
where $a_i$ is the cosmological scale factor at the redshift of lens
plane $i$ and ${\vct x}_{ik}$ is the location of the $k$th lens (of mass
$m_{ik}$) on lens plane $i$.
Note that ${\vct x}_{ik}$ is the position projected onto the observer plane,
i.e., ${\vct x}_{ik} = \chi_{so} {\vct\theta}_{ik}$ where ${\vct\theta}_{ik}$
is the angular position of the lens from the source.
Henceforth, we shall assume for simplicity that all the point masses have
the {\it same} mass $m$.
Then
\begin{equation}
{\vct s}_i ({\vct x}_i) = -x^2_{E,i}
  \sum_k {{\vct x}_i-{\vct x}_{ik} \over |{\vct x}_i-{\vct x}_{ik}|^2},
\label{eq5}
\end{equation}
where
\begin{equation}
x_{E,i} \equiv \left(4Gm \chi_{io}\chi_{so}\over a_i\chi_{si}\right)^{1/2}
\label{eq6}
\end{equation}
is the (Lagrangian) Einstein radius of a point mass on the $i$th lens plane.
For a random distribution of point masses, the optical depth to microlensing
is $\tau = \pi \sum_{i=1}^N n_i x_{E,i}^2$,
where $n_i = (a_i \chi_{si}/\chi_{so})^2 \Sigma_i/m$ and $\Sigma_i$ are,
respectively, the projected (onto the observer plane) surface number
density and the physical surface mass density of point masses on lens plane
$i$.
When we consider the situation where a significant fraction of the dark
matter in the universe is in the form of cosmologically distributed compact
objects, since the mean of the density inhomogeneities should be zero,
we should compensate the randomly distributed point masses on each plane by
a {\it negative} smooth surface mass density.
Then a term of the form $\pi n_i x_{E,i}^2 {\vct x}_i$ should be added to
the right hand side of equation (\ref{eq5}).

\section{MULTIPLANE LENSING BY A FEW POINT MASSES}

As we mentioned in \S~1, numerical simulations of gravitational microlensing
by a three-dimensional random distribution of point masses have revealed
differences between two- and three-dimensional microlensing (Refsdal 1970;
Schneider \& Weiss 1988a,b; Rauch 1991).
To find out and understand qualitatively some of these differences,
it is useful to study in detail specific cases of simple lens configurations.
In this section we study numerically the properties of multiplane
gravitational lensing by a few point masses.
To reduce the number of parameters,
we restrict our study to cases where the distance between the source and the
observer is much smaller than the horizon scale, i.e., $a \approx a_0$.

If there are $N_i$ point masses on lens plane $i$ (we shall denote the
total number of point masses as $N_\ast = \sum_{i=1}^N N_i$),
the sum of the squared (Lagrangian) Einstein radii is
\begin{equation}
x_E^2 \equiv \sum_{i=1}^N N_i x_{E,i}^2 .
\label{eq7}
\end{equation}
For the numerical computations, it is convenient to express all positions in
units of $x_E$ and to define a dimensionless mass parameter
\begin{equation}
m_i' \equiv {x_{E,i}^2 \over x_E^2} ,
\end{equation}
so that $\sum_{i=1}^N N_i m_i' = 1$.
Then the multiplane lens equation with identical point masses (eqs.[\ref{eq1}]
and [\ref{eq5}]) can be written as
\begin{equation}
{\vct x}_j = {\vct x}_1 - \sum_{i=1}^{j-1} \beta_{ij} m_i' \sum_k
             {{\vct x}_i-{\vct x}_{ik} \over |{\vct x}_i-{\vct x}_{ik}|^2} .
\end{equation}

\subsection{Configurations with Two Point Masses}

First let us consider in detail configurations with two point masses.
Without any loss of generality, we can choose a coordinate system such that
${\vct x}_{11} = 0$ and ${\vct x}_{21} = (d,0)$.
Since we assume that $a \approx a_0$, the two remaining parameters,
$\beta_{12}$ and $m_1'$, are related by $m_1' = 1/(2-\beta_{12})$.

The topology of the critical curves (and the caustics) as a function of
the parameters $(d, \beta_{12}, m_1')$ has been classified by ES (see
also Schneider \& Weiss 1986 for the $\beta_{12} = 0$ case).
In general, for given $\beta_{12}$ and $m_1'$, there are five transitions
in the topology of the critical curves as $d$ increases from $0$ (see Fig.~6
of ES).
However, in the limit $\beta_{12} \rightarrow 0$ (i.e., in the limit of two
point masses on the same plane), three of the transitions merge at $d = 0$
and the number of transitions is reduced to two.
For comparison, we show in Figures 2--7 two sets of models with
$\beta_{12} = 0$ and $\beta_{12} = 0.3$.
(We have also computed models with $\beta_{12} = 0.6$, but the results
are qualitatively similar to those with $\beta_{12} = 0.3$ and are not
shown here.)
The configurations shown in Figures 2--7 were chosen to sample regions of the
parameter space with different critical curve topologies.
The parameters $(\beta_{12}, d)$ and the critical curve topology types (in
the notation of ES) of these configurations are listed in Table 1.
(Note that the topology type $F$ is not represented;
with $m_1' = 1/(2-\beta_{12})$, type $F$ only occurs at very small $d$.)

In Figures 2 and 5 we show the critical curves of the configurations with
$\beta_{12} = 0$ and $0.3$ respectively.
The corresponding caustics and iso-magnification contours (on the observer
plane) are shown in the left and right panels of Figures 3 and 6.
In Figures 4 and 7 we plot the ``normalized'' differential cross-section
$\varphi(A) \equiv \sigma(A)/\sigma_0 (A)$, where $\sigma(A)$ is the
differential cross-section for magnification $A$ and
\begin{equation}
\sigma_0(A) = {2\pi x_E^2 \over (A^2 - 1)^{3/2}}
\end{equation}
is the differential cross-section of a single isolated point mass lens
with Einstein radius $x_E$ as defined in equation (\ref{eq7}).
As we saw in Paper I (see also \S~4 below), the function $\varphi(A)$ is a
useful measure of the deviations of the cross-section of a more complex lens
configuration from that of the point mass lens.
The cross-sections and the iso-magnification contours shown in these (and
all subsequent) Figures were obtained using the ray-shooting method (see,
e.g., Kayser, Refsdal, \& Stabell 1986; Schneider \& Weiss 1986, 1988b).
Since the magnification factor is calculated in this method by collecting
rays into a grid of square pixels, it is the magnification for an observing
device (or equivalently a source) of the size of one pixel.
To assess the effects of finite pixel size, we performed two sets of
calculations, with the linear pixel sizes in one set a factor $1.5$--$2$
smaller (see Table 1).
The pixel sizes were chosen so that all (except the very small) caustics
are resolved.
We show in Figures 3 and 6 the contours from the calculations with the
smaller pixels and in Figures 4 and 7 the function $\varphi(A)$ from both
sets of calculations.
As expected, the finite pixel size introduces an upper limit to the
magnification factor and produces extra ``bumps'' in $\varphi(A)$
(Rauch \etal\ 1992), but it does not have significant effects on the position
or the strength of the real caustic-induced features in $\varphi(A)$.

The $\beta_{12} = 0$ (and $m_1' = 0.5$) models have already been
studied in detail by Schneider \& Weiss (1986);
hence, we only give a brief description of these models here.
For $d \gg 2$, the region near each of the point masses is perturbed by the
weak shear from the other point mass.
So we have two critical lines that are slightly flattened ellipses and two
caustics that have the familiar ``astroid'' shape (see \S~3 and Fig.~2 of
Paper I).
As $d \rightarrow 2$, the critical curves and the caustics become asymmetric
(Figs.~2{\it a} and 3{\it a}).
The corresponding differential cross-section (Fig.~4{\it a}) shows
caustic-induced features that are similar to those found in the case of a
point mass perturbed by shear (see Fig.~3 of Paper I).
There is, in addition, a bump (at $A \approx 2$) which is associated with
the last iso-magnification contour that encloses both of the caustics.
As $d$ decreases below $2$, the number of critical curves (and caustics)
changes from two to one to finally three,
and the topology of the critical curves (and the caustics) changes from
$A'$ to $B'$ to finally $C'$ (see Table 1).
In all cases, there are caustic-induced features in the differential
cross-section at moderate to high $A$ and the cross-section at small $A$ is
reduced.
Each of these features is associated with either a local minimum or saddle
point in the magnification pattern or an iso-magnification contour that
touches the caustics at one or more points.
In the latter case, the contours of slightly larger $A$ are broken into
disconnected shorter segments by the caustics and the loss of the area
inside the caustics results in a drop in $\varphi(A)$.
(An example that has been discussed in detail in \S~3 of Paper I is the
feature associated with the contour osculating the astroid-shaped caustic
of the point mass plus weak shear lens.)

When the two point masses are not on the same lens plane,
there are new features and transitions in the critical curves and the
caustics (see Figs.~5 and 6).
At large Lagrangian separation $d$ (Fig.~5{\it a}),
besides the two critical curves that are similar to the ones seen in
the $\beta_{12} = 0$ case (see \S~4 for an explanation),
there is an extra critical curve inside the critical curve that
encloses point mass $1$ (i.e., the point mass closer to the source).
This critical curve is offset from point mass $1$ in the direction opposite
point mass $2$ and both its size and its distance from point mass $1$
increase with decreasing $d$.
The corresponding caustic is located at a large distance along the positive
$x$-axis (Fig.~6{\it a}).
This caustic has the (asymmetric) astroid shape of the two caustics that
are known from the $\beta_{12} = 0$ case, and it increases in size and
moves toward the other two caustics as $d$ decreases.
The critical curve (and caustic) configuration at large $d$ that we have
just described is denoted as topology type $A$ by ES.
Except for the presence of the new critical curve and caustic,
the topological transitions shown in Figures 5{\it a}--{\it c} and
6{\it a}--{\it c} are similar to those shown in Figures 2 and 3.
The remaining transitions do not exist in the single lens plane case.
However, as ES pointed out, except for the elliptic umbilic transition
between $C$ and $D$, all transitions are the beak-to-beak type.

We can explain the new critical line (and caustic) and its qualitative
behaviors if we focus on the configurations of topology types $A$--$C$.
For these configurations, it is clear that the new critical line is
composed of the set of rays that are strongly deflected by point mass $1$
(on lens plane $1$) to the vicinity of point mass $2$ (on lens plane $2$)
and are then infinitely magnified by point mass $2$.
Obviously this process can only occur when the point masses are not on the
same lens plane.
The new critical line is closer to point mass $1$ for larger $d$ because
a light ray must be more strongly deflected by point mass $1$ to reach
the vicinity of point mass $2$.

In the magnification patterns (Fig.~6), there are local minima and saddle
points that are associated with the new caustic;
there are also iso-magnification contours that touch the new caustic at two
points (e.g., the second outermost contour in Fig.~6{\it d}).
These new features in the magnification patterns are expected to produce
corresponding features in the cross-sections.
However, as we can see in Figure 7, the ``normalized'' differential
cross-sections $\varphi(A)$ are actually quite similar to those shown in
Figure 4.
Features associated with the new caustic are not noticeable until
$d \la 0.6$, and even in cases where the new caustic (which is
transformed at small $d$) encloses a large area of the observer plane,
the new features are typically weak and at $A \la 2$.
(They are stronger and at higher $A$ for the $\beta_{12} = 0.6$ models.)

If we examine the magnification patterns (and Fig.~8) more carefully,
we see that the new caustic is usually located in regions where the
background magnification is small.
In addition, the magnification throughout most of the region inside the new
caustic and, in particular, the minimum magnification inside the new caustic
are only slightly higher than the low background magnification.
Hence the new features in the magnification patterns and the corresponding
features in $\varphi(A)$ have small $A$ values.
Next, let us consider specifically the feature associated with the
iso-magnification contour that touches the new caustic at two points
(e.g., the second outermost contour in Fig.~6{\it d}).
This contour is long and, except for the distortion near the new caustic,
nearly circular.
Contours of slightly higher magnification are broken by the new caustic
into a short arc around one of the cusps and a long arc that encloses the
other caustic(s).
This results in a decrease in the cross-section $\sigma(A)$ because an area
that would otherwise be enclosed by two of these contours (of magnification
$A$ and $A+dA$) is lost to the interior of the caustic.
Although the area lost to the caustic can be large, it is typically much
smaller than the area enclosed by the long (and the short) arcs.
Consequently, the decrease in $\varphi(A) = \sigma(A)/\sigma_0(A)$, which is
approximately the fractional decrease in $\sigma(A)$, is small.
Similarly, the increase in $\varphi(A)$ at $A = A_{\rm min}$, where
$A_{\rm min}$ is the minimum magnification inside the new caustic, is small
because $A_{\rm min}$ is typically small and the area enclosed by the
contours of magnification $A_{\rm min}$ and $A_{\rm min} + dA$ outside the
new caustic (mainly two long arcs) is much larger than the area enclosed by
the contour $A_{\rm min} + dA$ inside the new caustic.

The relatively weak influence of the new caustic on the magnification
pattern is demonstrated more clearly in the sample of magnification profiles
plotted in Figure 8.
Although the magnification must be infinite on the caustic,
only a very small region near the caustic has $A \gg 1$.
In fact, the numerically obtained $A$, which is finite at a caustic because
of the finite pixel size of our calculations, shows a maximum at the new
caustic that is quite small.
This behavior of the magnification pattern near the new caustic matches the
description of the ``thin caustics'' found in the microlensing simulations
of Schneider \& Weiss (1988b).
We have therefore found in the simple two-point-mass case a clear
demonstration of the occurrence of thin caustics.
As we mentioned earlier, the new caustic consists of light rays that
are strongly deflected by the first point mass to the vicinity of the second
point mass and are then infinitely magnified by the second point mass.
In the simulations of Schneider \& Weiss, which have a large number of point
masses distributed on several lens planes, a light ray can be strongly
deflected on more than one lens plane.
It is likely that most of the caustics that are strongly deflected on one or
more lens planes (i.e., the secondary caustics defined in \S~3.2 below) are
thin caustics.
This identification for the thin caustics is consistent with the conjecture
by Schneider \& Weiss that the thin caustics are caused by steep deflection
gradients because the regions (in Lagrangian space) with strong deflections
are also regions with steep deflection gradients.

\subsection{Configurations with Three and Four Point Masses}

Now let us consider briefly configurations with three and four point masses.
Two configurations with $N_\ast = 3$ point masses and one with $N_\ast = 4$
are shown in Figures 9--11.
These configurations have one point mass on each lens plane (i.e.,
$N = N_\ast$) and their parameters are listed in Table 2.
The configuration {\it a} with $N_\ast =3$ was chosen because it has the
critical curve topology of point masses widely separated in Lagrangian
space while the smaller critical curves are clearly resolved.
The other configurations are ``typical'' configurations resulting from
randomly distributing the point masses within a cone of constant Lagrangian
radius.
We have (arbitrarily) chosen a cone radius such that a light ray passing
within the cone would be infinitely magnified if the mass of the lenses
were uniformly distributed within the cone.

A striking feature of Figure 9 (and Fig.~5) is the rapid increase in the
number of critical curves with $N_\ast$.
For a configuration with $N_\ast$ point masses well separated in Lagrangian
space and with each point mass on its own lens plane (i.e., $N = N_\ast$),
we can calculate the number of critical curves by extending the discussion
given in \S~3.1 for the origin of the new critical curve.
Light rays from the source can reach the vicinity of the point mass on the
$i$th lens plane after being strongly deflected by any combination (or none)
of the previous $i-1$ point masses.
For each combination, there is a set of rays that are infinitely magnified
by the $i$th point mass (and the weak deflections on the remaining planes).
Therefore, there are $2^{i-1}$ critical curves associated with the $i$th
point mass and the total number of critical curves is $2^{N_\ast} -1$.
The critical curves shown in Figures 5{\it a} and 9{\it a} are consistent
with these numbers.
(The total number of critical curves is smaller for configurations with
$N < N_\ast$ and approaches $N_\ast$ in the limit $N = 1$, i.e., in the
single lens plane case.)
Hereafter, we shall refer to the caustics that are not strongly deflected
as the {\it primary} caustics and those that are strongly deflected on at
least one lens plane as the {\it secondary} caustics.

Although the number of critical curves increases exponentially with $N_\ast$,
we can see in Figure 9 that the size of a critical curve decreases
rapidly as the number of strong deflections associated with the critical
curve (or equivalently the number of critical curves within which it is
nested) increases.
On the other hand, as in the two-point-mass case, a caustic formed after
a larger number of strong deflections is not necessarily smaller.
For example, in Figure 10{\it c}, the caustic located near $(-0.3,1.2)$
[which corresponds to the critical curve located near $(0.25,-0.4)$]
and the caustic with six cusps [which corresponds to the largest critical
curve] are comparable in size.

In the two-point-mass case, due to the absence of overlapping or
self-intersecting caustics,
a complete classification of the critical curve topology also provides
a complete classification of the caustic topology (ES).
This is not true when $N_\ast > 2$.
In particular, there are overlaps of caustics that belong to different
critical curves.
Examples of such overlapping caustics are the caustics located near
$(0.3,0.3)$ in Figure 10{\it a} and those in the lower left corner of
Figure 10{\it c}.
(There are no self-intersecting caustics in the configurations shown in
Fig.~10, but they may be present in other configurations.)

Finally, we discuss the normalized differential cross-sections plotted in
Figure 11.
For the ``typical'' configurations {\it b} and {\it c}, the qualitative
behaviors of $\varphi(A)$ agree with what one might expect from the
two-point-mass configurations studied in \S~3.1.
The cross-section at small $A$ is smaller than that for a single point mass.
There are prominent features at moderate to high $A$ that are associated
with the primary caustics.
Features associated with the secondary caustics are typically weak,
but some of them are noticeable at $A \la 2$.
For the configuration {\it a}, the features associated with the primary
caustics, i.e., those at $\log A \ga 0.8$, are similar to the features
in Figures 4{\it a} and 7{\it a}.
This is expected since configuration {\it a} has the critical curve
topology of widely separated point masses.
However, the features associated with the secondary caustics are at
unusually high $A$:
the ``plateau'' between $\log A \approx 0.55$ and $0.8$ consists of a
series of such features.
This is due to the fact that the parameters of configuration {\it a} were
specially chosen so that the smaller critical curves are clearly resolved.
For these parameters, the caustics are unusually close to one
another (see Fig.~10{\it a} and notice the change in scale between
Figs.~10{\it a} and 10{\it b}).
As a result, the secondary caustics are located in regions of relatively
high background magnification and the associated features in $\varphi(A)$
are at unusually high $A$.

\section{THREE-DIMENSIONAL LOW OPTICAL DEPTH MICROLENSING}

The study of simple lens configurations in the previous section shows
that microlensing by a three-dimensional distribution of point masses is,
in general, much more complicated than microlensing by lenses confined to
a single lens plane.
But there are also indications that, in the low optical depth limit, some
aspects of the three-dimensional problem can be reduced to the planar one.
For example, the multiplane configurations with point masses widely
separated in Lagrangian space show primary caustics and corresponding
features in $\varphi(A)$ that are similar to those found in the
single plane configurations.
In \S~4.1 we confirm this analytically for the limiting case of a point
mass lens perturbed by the weak shear from the other lenses.
Another useful result from the study in \S~3 is that the significant
features in $\varphi(A)$ at moderate to high $A$ are associated with the
primary caustics.
Therefore, we can neglect the features associated with the secondary
caustics when we analyze the macroimage magnification distribution,
$P(A)$, at high $A$.
In \S~4.2 we derive the feature induced by the primary caustics at
the high magnification end of $P(A)$ for a three-dimensional low optical
depth lens distribution.
The resulting semi-analytic $P(A)$ is compared to previous numerical
results in \S~4.3, and the differences between the two- and
three-dimensional results are discussed in \S~4.4.

\subsection{Cross-Section of the Point Mass Lens Plus Weak Shear in\\
            Three Dimensions}

In the low optical depth limit, lensing (other than low magnification events)
by a three-dimensional distribution of point masses is, as in the planar case,
typically dominated by a single point mass.
Without loss of generality, we shall designate one of the point masses
residing on the $L$th lens plane as the dominant one and locate this point
mass at the origin of the lens screen.
On all of the other screens, the light ray is only weakly perturbed, and
the perturbations are expanded to the first order in $\vct x_i$.
Before the light ray passes the $L$th screen with the dominant lens (i.e.,
for $j \leq L$), the lens mapping (eqs.[\ref{eq1}] and [\ref{eq5}]) is
simplified to
\begin{equation}
{\vct x}_j = {\vct x}_1 + \sum_{i=1}^{j-1}
             \beta_{ij} \left({\vct \alpha}_i + \shear_i {\vct x}_i\right),
\label{eq11}
\end{equation}
where the constant deflection
\begin{mathletters}
\begin{equation}
{\vct \alpha}_i = x^2_{E,i}
                  \sum_{k} {{\vct x}_{ik}\over |{\vct x}_{ik}|^2}
\end{equation}
and the shear matrix
\begin{equation}
\shear_i = x^2_{E,i} \sum_{k} {1 \over |{\vct x}_{ik}|^4}
           \left(\begin{array}{cc}
           x_{ik}^2 - y_{ik}^2&2 x_{ik} y_{ik}\\
           2 x_{ik} y_{ik}&y_{ik}^2 - x_{ik}^2
           \end{array}\right)
\label{eq12b}
\end{equation}
\end{mathletters}
(similar to ${\vct\alpha}_L$ and $\shear_L$ defined in eq.[13] of Paper I)
represent the perturbative influence of the lenses on the $i$th plane.
Equation (\ref{eq11}) can also be written as
\begin{equation}
{\vct x}_j = {\matrixmark B}_j {\vct x}_1 + {\vct u}_j,
\end{equation}
where
\begin{equation}
{\matrixmark B}_j = {\matrixmark I} + \sum_{i=1}^{j-1} 
                    \beta_{ij} \shear_i {\matrixmark B}_i ,
\qquad
{\vct u}_j = \sum_{i=1}^{j-1} \beta_{ij}
             \left({\vct \alpha}_i + \shear_i {\vct u}_i\right) ,
\end{equation}
and ${\matrixmark I}$ is the identity matrix.
Once the light ray has traversed the $L$th lens screen,
the location at which the light ray intersects subsequent (i.e., $j>L$)
screens is given by
\begin{equation}
{\vct x}_j = {\matrixmark B}_j {\vct x}_1 + {\vct u}_j +
             \beta_{Lj} {\matrixmark C}_j {\vct d}_L({\vct x}_L),
\end{equation}
where
\begin{equation}
{\vct d}_L ({\vct x}_L) = - x^2_{E,L} {{\vct x}_L\over |{\vct x}_L|^2}
\end{equation}
denotes the influence of the dominant lens and
\begin{equation}
{\matrixmark C}_j = {\matrixmark I} + \sum_{i=L+1}^{j-1}
             \left(\beta_{ij}\beta_{Li} \over \beta_{Lj}\right)
             \shear_i {\matrixmark C}_i.
\end{equation}
Note that, in the absence of the dominant lens, ${\vct u}_j$ is the path
followed by a light ray with ${\vct x}_1 = 0$.

On the observer plane [which is the $(N+1)$th screen, but we also denote
it with the subscript ``o''], the Eulerian position is given by
\begin{equation}
\vct r = {\matrixmark B}_o {\vct x} + {\vct u}_o +
         {\matrixmark C}_o {\vct d}_L({\vct x}_L).
\label{eq18}
\end{equation}
If we use the fact that ${\vct x} = {\vct x}_1 = {\matrixmark B}_L^{-1}
\left({\vct x}_L - {\vct u}_L\right)$ and transform the Eulerian variable
according to $\vct r \to {\vct r}' = {\matrixmark C}_o^{-1}
\left(\vct r - {\vct u}_o + {\matrixmark B}_o {\matrixmark B}_L^{-1}
{\vct u}_L\right)$, the above equation can be cast into the form
\begin{mathletters}
\begin{equation}
{\vct r}' = {\vct x}' + {\vct s}'({\vct x}'),
\end{equation}
where ${\vct x}'\equiv{\vct x}_L$,
\begin{equation}
{\vct s}'({\vct x}') = {\vct d}_L({\vct x}') + \shear'_L {\vct x}',
\qquad {\rm and} \qquad
\shear'_L = {\matrixmark C}_o^{-1} {\matrixmark B}_o {\matrixmark B}_L^{-1} -
            {\matrixmark I}.
\end{equation}
\end{mathletters}
Although there is only shear on each lens screen,
the matrix $\shear'_L$ is in general a combination of shear and convergence
(i.e., it is symmetric but not traceless).
Thus equation (19) is formally equivalent to the single plane lens
equation for the point mass lens plus shear and smooth surface density
(compare with eq.[15] of Paper I for the point mass plus shear lens).
This is a specific example of the ``non-linear telescope'' (or generalized
quadrupole lens) discussed by Kovner (1987; see also Schneider \etal\ 1992).

We are interested in the case where the shear perturbations on all the lens
screens are weak.
To first order in $\shear_i$, the expressions for ${\matrixmark C}_o$,
${\matrixmark B}_j$, and $\shear'_L$ are
\begin{mathletters}
\begin{equation}
{\matrixmark C}_o \approx {\matrixmark I} + \sum_{i=L+1}^N
                          \beta_{Li} \shear_i,
\qquad\quad\qquad
{\matrixmark B}_j \approx {\matrixmark I} + \sum_{i=1}^{j-1}
                          \beta_{ij} \shear_i,
\end{equation}
and
\begin{equation}
\shear'_L
\approx \sum_{i=1}^L \left(\chi_{si}\chi_{Lo}\over
                           \chi_{sL}\chi_{io}\right) \shear_i +
        \sum_{i=L+1}^N \left(\chi_{sL}\chi_{io}\over
                             \chi_{si}\chi_{Lo}\right) \shear_i ,
\label{eq20b}
\end{equation}
\end{mathletters}
respectively.
Hence $\shear'_L$ is reduced to a shear matrix:
$\shear'_L \approx S'(\chi_{Lo}) \tshear(\phi'_L)$, where $S'(\chi_{Lo})$ is
the magnitude of the ``effective'' shear and
\begin{equation}
\tshear(\phi'_L) = \left(\begin{array}{lr}
                   \cos 2\phi'_L& \sin 2\phi'_L\\
                   \sin 2\phi'_L&-\cos 2\phi'_L
                   \end{array}\right) .
\end{equation}
Then equation (19) reduces to the single plane lens equation for the
point mass plus weak shear lens (eq.[15] of Paper I).

In \S\S~3.1--2 of Paper I, we studied in detail the properties of single
plane lensing by the point mass plus weak shear lens.
The results are applicable to the mapping from $\vct x'$ to $\vct r'$
(eq.[19]).
We can then infer the properties of the lens mapping from $\vct x$ to
$\vct r$ (eq.[\ref{eq18}]) by taking into account the effects of the
transformations $\vct x \to \vct x'$ and $\vct r \to \vct r'$.
First, it is easy to see that, to first order in $\shear_i$, both
transformations are a combination of translation and shear distortion and
are one-to-one.
Second, since the deformation tensor
\begin{equation}
{\matrixmark D} \equiv {\partial \vct r \over \partial \vct x}
         = \left(\partial \vct r \over \partial {\vct r}'\right)
           \left(\partial {\vct r}' \over \partial {\vct x}'\right)
           \left(\partial {\vct x}' \over \partial \vct x\right)
         = {\matrixmark C}_o \left(\partial{\vct r}' 
		  \over\partial{\vct x}'\right) {\matrixmark B}_L,
\end{equation}
there is a simple correspondence between the caustics (and the critical
curves) of the mappings (\ref{eq18}) and (19).
In addition, because the determinants $|{\matrixmark C}_o|$ and
$|{\matrixmark B}_L|$ are $1 + O\bigl(S'^2\bigr)$,
the magnification $|{\matrixmark D}|^{-1}$ at a point $\vct r$ is,
to first order in $S'(\chi_{Lo})$, identical to the ``magnification''
$|\partial{\vct r}'/\partial{\vct x}'|^{-1}$ at the corresponding point
${\vct r}'$.
Similarly, the correction factors for the differential cross-section are
$1 + O\bigl(S'^2\bigr)$.
All of the corrections that we have just considered are of the same order
as the terms already ignored in the weak shear analysis in Paper I.
Therefore, the effects of the transformations are negligible,
and the properties derived in \S\S~3.1-2 of Paper I are also applicable to
the lens mapping (\ref{eq18}).
Briefly, for a point mass lens perturbed by weak shear in three dimensions,
the critical line is a slightly flattened ellipse (of size $x_{E,L}$);
the caustic has the shape of an astroid, with the four cusp catastrophes (at
a distance $2 x_{E,L} S'(\chi_{Lo})$ from the center of the astroid)
connected to each other by four fold catastrophes (see Fig.~2 of Paper I);
the differential cross-section has the scaling property (see eq.[23] of
Paper I)
\begin{equation}
\sigma(A|S'[\chi_{Lo}]) = {2 \pi x_{E,L}^2 \over (A^2 -1)^{3/2}} \,
                          \varphi(A S'[\chi_{Lo}]),
\label{eq23}
\end{equation}
where the normalized differential cross-section $\varphi(A S')$ depends on
$A$ and $S'(\chi_{Lo})$ only through the combination $A S'$;
and $\varphi(A S')$, which is shown in Figure 3 of Paper I, shows strong
caustic-induced features at the critical values $A_1 = 2/3\sqrt{3}S'$ and
$A_2 = 1/S'$.

As we mentioned in \S~2, when we consider the situation where a significant
fraction of the dark matter in the universe is in the form of cosmologically
distributed compact objects,
we should compensate the randomly distributed point masses on each lens
plane by a negative smooth surface mass density.
The derivation in this section can be generalized to this case if we replace
$\shear_i$ in all the equations (except eq.[\ref{eq12b}]) by
$\shear_i + \pi n_i x_{E,i}^2 {\matrixmark I}$.
The results remain valid, but the magnification $A$ should be interpreted
as the magnification with respect to an ``empty beam''
(i.e., a light beam that is not affected by the point masses).

\subsection{Three-Dimensional Superposition of Cross-Sections
            with Distributed Shear}

As we discussed earlier, in the case of low optical depth microlensing by a
three-dimensional distribution of point masses,
features associated with the secondary caustics are negligible at the high
magnification end of the macroimage magnification distribution $P(A)$.
Therefore, as in the two-dimensional case (see \S~3.3 of Paper I),
$P(A)$ at high $A$ can be approximated as a superposition of the
cross-sections of the individual point mass lenses,
with the shear perturbation due to the other point masses varying from
lens to lens.

Let us assume that the point masses are distributed randomly on $N$ lens
planes, each with projected (onto the observer plane) surface number density
$n_i$.
Then $P(A)$ can be written as
\begin{equation}
P(A) = n \sum_{L=1}^N p(\chi_{Lo})
       \int_0^\infty dS' \, p(S'|\chi_{Lo}) \, \sigma(A|S') ,
\label{eq24}
\end{equation}
where $\sigma(A|S')$ is the cross-section given in equation (\ref{eq23}),
$p(S'|\chi_{Lo})$ is the probability distribution of the effective shear
$S'(\chi_{Lo})$ for a dominant lens at $\chi_{Lo}$, $p(\chi_{Lo})$ is the
probability that the dominant lens is located on the $L$th screen,
and $n = \sum_{i=1}^N n_i$ is the total projected surface number density.
Since $p(\chi_{Lo})$ is simply
\begin{equation}
p(\chi_{Lo}) = {n_L \over n} ,
\end{equation}
we have
\begin{equation}
P(A) = \sum_{L=1}^N P(A|\chi_{Lo}) ,
\label{eq26}
\end{equation}
where
\begin{equation}
P(A|\chi_{Lo})
     \equiv n_L \int_0^\infty dS' \, p(S'|\chi_{Lo}) \, \sigma(A|S') .
\label{eq27}
\end{equation}
The function $P(A|\chi_{Lo})$ has the same structure as $P(A)$ of the
two-dimensional case (eq.[29] of Paper I), and it is the contribution to the
macroimage magnification distribution by the lenses on the $L$th lens screen
as the dominant lenses.

The effective shear $S'(\chi_{Lo})$ that enters into the cross-section
$\sigma(A|S')$ is a weighted superposition of the shear induced by all except
the dominant lens (eq.[\ref{eq20b}]).
In the Appendix we show that
\begin{equation}
p(S'|\chi_{Lo}) =
{\tau'(\chi_{Lo}) S' \over \left[\tau'^2(\chi_{Lo}) + S'^2\right]^{3/2}},
\label{eq28}
\end{equation}
where
\begin{equation}
\tau'(\chi_{Lo}) = \sum_{i=1}^{L}
    \left(\chi_{si}\chi_{Lo}\over\chi_{sL}\chi_{io}\right) \pi n_i x^2_{E,i} +
                   \sum_{i=L+1}^N
    \left(\chi_{sL}\chi_{io}\over\chi_{si}\chi_{Lo}\right) \pi n_i x^2_{E,i}
\end{equation}
is the ``effective'' optical depth at $\chi_{Lo}$.
[Note that the effective optical depth $\tau'(\chi_{Lo})$ is smaller than
the (usual) optical depth $\tau = \sum_{i=1}^N \pi n_i x_{E,i}^2$ because
the extra factors $(\chi_{si}\chi_{Lo}/\chi_{sL}\chi_{io})$ (for $i < L$)
and $(\chi_{sL}\chi_{io}/\chi_{si}\chi_{Lo})$ (for $i > L$) are less than
unity.]
Substituting equations (\ref{eq23}) and (\ref{eq28}) into equation
(\ref{eq27}),
we find that $P(A|\chi_{Lo})$ is similar to $P(A)$ for a single lens plane
(compare with eq.[31] of Paper I):
\begin{equation}
P(A|\chi_{Lo}) = {2 \pi n_L x_{E,L}^2 \over (A^2 - 1)^{3/2}}
                 f_1 [A \tau'(\chi_{Lo})],
\end{equation}
where $f_1(A \tau') = \int_0^\infty dy\,\varphi(A \tau' y) y/(1+y^2)^{3/2}$
is the function introduced in \S~3.3 of Paper I to describe the
caustic-induced feature in $P(A)$ in the single lens plane case (see Fig.~4
of Paper I; also Fig.~12 of this paper).
Thus $P(A|\chi_{Lo})$ exhibits the same caustic-induced bump (with a maximum
enhancement of $20\%$ at $A \approx 2/\tau'$) and the same scaling property
(with $f_1$ being a function of the combination $A \tau'$) as $P(A)$ of the
single lens plane case.

There is, however, a significant difference between the two- and
three-dimensional cases.
To obtain the macroimage magnification distribution in the
three-dimensional case,
we must sum $P(A|\chi_{Lo})$ over all the lens planes (eq.[\ref{eq26}]):
\begin{equation}
P(A) = \sum_{L=1}^N {2 \pi n_L x_{E,L}^2 \over (A^2 - 1)^{3/2}}
                     f_1[A \tau'(\chi_{Lo})]
     = {2 \tau \over (A^2 - 1)^{3/2}} f_N (A; \tau) ,
\label{eq31}
\end{equation}
where
\begin{equation}
f_N (A; \tau) \equiv \sum_{L=1}^N \left(\pi n_L x_{E,L}^2 \over \tau\right)
                     f_1 \left[A \tau {\tau'(\chi_{Lo})\over \tau}\right] ,
\label{eq32}
\end{equation}
and $\tau = \pi \sum_{L=1}^N n_L x_{E,L}^2$ is the (usual) optical depth.
Although we denote the caustic-induced feature in the $N$-plane case as
$f_N (A; \tau)$ for simplicity,
it should be clear from equation (\ref{eq32}) that $f_N (A; \tau)$ depends
on the location and the surface density of the lens planes in a non-trivial
manner and does not have any simple scaling properties (when $N > 1$).
Note also that the differences between $f_1 (A\tau)$ and $f_N (A; \tau)$
are due to the differences between the optical depths $\tau$ and
$\tau'(\chi_{Lo})$.

To proceed further, we need to assume a model for the distribution of the
point mass lenses.
As in the numerical simulations reported by several authors (Refsdal 1970;
Schneider \& Weiss 1988a,b; Rauch 1991),
we consider a lens distribution with {\it constant} comoving mass density
$\rho_l$.
We divide the volume between the source and the observer into $N$ shells
of equal comoving thickness ($\Delta\chi = \chi_{so}/N$) and place the point
masses in each shell on a lens plane at the middle of the shell.
In Figure 12 we plot the numerically evaluated $f_N (A; \tau)$ for $N = 1$,
$2$, $4$, $8$, and $16$, and for source redshift $z_s \to 0$ and $= 2$.
It is clear that $f_N$ has converged by $N = 16$.
This agrees with the conclusion from a study of lensing by smoothly variable
mass distribution that a fully three-dimensional mass distribution is well
approximated by as few as eight or $16$ screens (Lee \& Paczy\'nski 1990).
It is also clear that $f_\infty$($\equiv \lim_{N\to\infty} f_N$) is very
insensitive to the source redshift, at least for $z_s \la 2$.

In the limit $N \to \infty$, the summations in the expressions for $\tau$,
$\tau'(\chi_{Lo})$, and $f_N (A; \tau)$ are replaced by integrals:
\begin{mathletters}
\begin{eqnarray}
\tau &=& 4 \pi G \rho_l a_0^3 \int_0^{\chi_{so}} d\chi
         {\chi \left(\chi_{so}-\chi\right) \over a(\chi) \chi_{so}} ; \\
\tau'(\chi_{Lo}) &=& 4 \pi G \rho_l a_0^3 \left[
                     \int_0^{\chi_{sL}} d\chi {\chi_{Lo} \chi^2 \over
                                               a(\chi) \chi_{so} \chi_{sL}} +
                     \int_{\chi_{sL}}^{\chi_{so}} d\chi
                         {\chi_{sL} \left(\chi_{so}-\chi\right)^2 \over
                          a(\chi) \chi_{so} \chi_{Lo}}
                     \right] ; \\
f_\infty (A; \tau) &=& {4 \pi G \rho_l a_0^3 \over \tau}
                   \int_0^{\chi_{so}} d\chi
                   {\chi \left(\chi_{so}-\chi\right) \over a(\chi) \chi_{so}}
                   f_1 \left[A \tau {\tau' \over \tau}\right] .
\end{eqnarray}
\end{mathletters}
Equations (33a--b) can be evaluated analytically, and they are
particularly simple in the limit $z_s \to 0$ (i.e., in the limit
$a(\chi_{so}) \approx a_0$):
\begin{mathletters}
\begin{eqnarray}
\tau &\approx& {2 \pi \over 3} G \rho_l a_0^2 \chi_{so}^2 ; \\
\tau'(\chi_{Lo}) &\approx& {4 \pi \over 3}
                           G \rho_l a_0^2 \chi_{sL} \chi_{Lo} ; \\
f_\infty (A; \tau) &\approx& 6 \int_0^{\chi_{so}} d\chi
                   {\chi \left(\chi_{so}-\chi\right) \over \chi_{so}^3}
                   f_1 \left[A \tau {2 \chi (\chi_{so}-\chi) \over
                             \chi_{so}^2}\right] .
\end{eqnarray}
\end{mathletters}
Thus the contribution to $f_\infty$ is dominated by the lens planes at
$\chi_{Lo} \approx \chi_{so}/2$ with $\tau'/\tau \la 1/2$.
This is consistent with the numerical results shown in Figure 12.
The most significant change from $f_1$ to $f_\infty$ is the shift in the
location of the bump from $A \tau \approx 2$ to $A \tau \approx 5$.
Otherwise, the peak value ($\approx 18\%$ enhancement) and the width of
$f_\infty$ are only slightly smaller and wider.

\subsection{Comparison with Numerical Simulations}

Several authors (Refsdal 1970; Schneider \& Weiss 1988a,b; Rauch 1991) have
used numerical simulations to calculate the macroimage magnification
distribution, $P(A)$, produced by a three-dimensional distribution of point
masses.
In the most recent study by Rauch, distributions with high resolution in
$A$ were obtained from Monte Carlo simulations for three cases with
$\sigma = 0.01$, $0.1$, and $0.2$.
The ``equivalent $\sigma$ value'' is related to the average magnification
(with respect to an empty beam) on the observer plane, $\bar A$, by
$\bar A = (1 - \sigma)^{-2}$, i.e., $\sigma$ is the optical depth $\tau$ of
a {\it two-dimensional} lens distribution with the same $\bar A$.
For a three-dimensional lens distribution, $\sigma$ and $\tau$ are not
identical in general, but $\bar A \approx 1 + 2\tau \approx 1 + 2\sigma$
in the limit $\tau \ll 1$.

We wish to compare the semi-analytic $P(A)$ derived in \S~4.2 to the
distributions obtained by Rauch.
(We shall focus on his results for an Einstein-de Sitter cosmological
background, but Rauch also found that the distributions for the same $\sigma$
but different background cosmology are nearly identical.)
Since our derivation assumes low optical depth,
the best case for comparison should be the $\sigma = 0.01$ case.
Unfortunately, even with $5 \times 10^5$ simulations, the Monte Carlo
simulations only provide useful information about $P(A)$ for $A < 50$
(see Fig.~2 of Rauch 1991).
This is not sufficient for detecting the caustic-induced feature because
significant deviation from $P(A) = 2\tau/(A^2-1)^{3/2}$ is not expected for
$A \la 1/\tau$ ($\approx 100$ for $\tau \approx \sigma = 0.01$)
(see $f_\infty$ in Fig.~12).
We shall not show the comparison for this case and simply note that the
numerically obtained $P(A)$ (for $A < 50$) agrees with the low optical
depth semi-analytic result, with or without the caustic-induced feature.

In Figure 13 we show the distributions obtained by Rauch for $\sigma = 0.1$
and $0.2$ (histograms).
The data are those shown in Figures 3 and 4 of Rauch (1991),
but they are plotted in the form $(A^2 - 1)^{3/2} P(A)$ and we have increased
the bin sizes at large $A$ to reduce statistical fluctuations.
As Rauch \etal\ (1992) pointed out, the distributions in the
three-dimensional case do not show the prominent bump at $A \ga 2.5$
that was found in the planar case.
However, the distributions are not featureless.
This can be seen most clearly in the $\sigma = 0.2$ case
(note, e.g., the inflection point at $\log A \approx 0.6$).

In a plot of $(A^2 - 1)^{3/2} P(A)$, the low optical depth distribution
without the caustic-induced feature is a horizontal line of amplitude
$K = 2\tau$, while the semi-analytic distribution with the caustic-induced
feature is $K f_\infty(A; \tau)$ (eq.[\ref{eq31}]; recall that $f_\infty$ is
very insensitive to the source redshift).
Since $\sigma = 0.1$ and $0.2$ are not much less than unity,
two corrections to the low optical depth semi-analytic results are necessary
when we compare them to the numerical ones.
First, it is no longer accurate to assume that $\tau = \sigma$.
For these particular models, $\sigma = 0.1$ and $0.2$ correspond to
$\tau = 0.108$ and $0.233$, respectively.
Second, the normalization $K$ can be significantly different from the low
optical depth result $2\tau$.
In \S~5.1 of Paper I, we discussed the renormalization of $P(A)$ due to the
net convergence of light rays and multiple streaming for the single lens
plane case.
 From an analytic result for $P(A)$ in the high magnification limit, which
is exact even for finite optical depth $\tau$ (see also Schneider 1987),
we adopted the normalization
\begin{equation}
K = 2 \tau {\bar A \over (1 + \tau^2)^{3/2}} .
\label{eq35}
\end{equation}
(For $\tau \ll 1$, $\bar A \approx 1 + 2\tau$ and $(1 + \tau^2)^{-3/2}
\approx 1 - 3\tau^2/2$, and the correction to the normalization $2\tau$ is
small.)
The analysis that led to equation (\ref{eq35}) cannot be easily generalized
to the three-dimensional case (it would require the statistical properties
of the matrix ${\matrixmark B}_o$ which depends on the shear matrices
$\shear_i$ {\it non-linearly}; see eq.[14]), and an equivalent expression
for $K$ for a three-dimensional lens distribution is not known.
Nevertheless, in the range of $\tau$ of interest here (i.e., for
$\tau \la 0.25$), it is likely that equation (\ref{eq35}) is a good
approximation for $K$ in the three-dimensional case.
In equation (\ref{eq35}), the dominant correction to the normalization $2\tau$
is the factor $\bar A$.
(The correction due to the factor $(1 + \tau^2)^{-3/2}$ is second order in
$\tau$ and less than $10\%$ for $\tau \la 0.25$.)
The factor $\bar A$ arises from converting a probability distribution in
Lagrangian space to a distribution in Eulerian space (see \S~5.1 of Paper I).
The conversion (and hence the factor $\bar A$) should remain in any
generalization of the analysis to the three-dimensional case.
Furthermore, as in the planar case (see \S~5.1 of Paper I), we can derive
the factor $\bar A$ using a different argument:
in \S~4.2 we assumed that the surface density of the astroid-shaped
caustics on the observer plane is the same as the projected surface density
of point masses $n$ (see eq.[\ref{eq24}]);
there is, however, a net convergence of the light rays by the overall lens
distribution;
this increases the density of astroids by a factor $\bar A$, and introduces
an additional factor $\bar A$ in the normalization of $P(A)$ in equation
(\ref{eq31}).
Therefore, we shall adopt equation (\ref{eq35}) for the normalization $K$ in
the three-dimensional case.

In Figure 13 we plot $K$ (dashed lines) and $K f_\infty(A; \tau)$ (solid
lines) for $\sigma = 0.1$ and $0.2$.
They should be compared to the numerical results at $\log A \ga 0.3$
(see \S~4.5 for a brief discussion of the deviations at small $A$).
As we can see, the adopted normalization $K$ is in good agreement with
the amplitudes of the numerical results.
In addition, the numerical results show deviations from the horizontal
dashed lines that are similar in shape to the semi-analytic caustic-induced
feature $f_\infty$
(note, e.g., that the inflection point in the numerical result
for $\sigma = 0.2$ coincides with the location $A \approx 1/\tau$ where
$f_\infty$ starts to deviate significantly from unity).
In the $\sigma = 0.1$ case, there is little, if any, difference between the
numerical and semi-analytic results;
the only possibly statistically significant deviation is that the numerical
distribution may be higher near $\log A \approx 1$.
In the $\sigma = 0.2$ case, the difference between the numerical and
semi-analytic results is more pronounced, with the numerical distribution
slightly higher than the semi-analytic distribution over the range
$0.7 < \log A < 2$.
Note, however, that the number of events in the histograms at $\log A > 1.6$
is small (see the fluctuations amongst the same bins in the $\sigma = 0.1$
case), and the true distribution may not be as flat as the histograms shown.

\subsection{Criterion for Low Optical Depth}

The comparison in \S~4.3 shows that, in the case of a three-dimensional lens
distribution, the function $f_\infty$, which was derived in the low optical
depth limit, provides a reasonably good fit to the caustic-induced feature
in $P(A)$ for $\tau$ as large as $0.233$.
This is very different from the two-dimensional case studied in Paper I.
For a two-dimensional lens distribution, the caustic-induced feature in the
numerically obtained $P(A)$ is more prominent than the low optical depth
semi-analytic feature $f_1$ for $\tau$ as small as $0.1$.
This was shown in Figure 6 of Paper I where we compared the numerical and
semi-analytic distributions for $\tau (= \sigma) = 0.1$ and $0.2$ in a plot
of $(A^2-1)^{3/2} P(A)$.
The numerical distributions show a clear minimum at $A \approx 2.5$
and are significantly higher than the semi-analytic results in the range
$0.5 \la \log A \la 1$.

For both the two- and three-dimensional lens distributions, the low optical
depth analysis assumes that the point mass lenses are well separated in
Lagrangian space.
Then the deflection near each point mass is due to that point mass and the
shear perturbation from the other lenses, and an astroid-shaped (primary)
caustic is associated with each point mass.
The low optical depth analysis is not valid if the lens density is high
enough that a significant fraction of the lenses have close neighbor(s),
and their caustics are complex structures produced by the collective
effect of two (or more) point masses.
In \S~5.3 of Paper I, we estimated for the two-dimensional case the
fraction of point masses whose caustics are not isolated astroids,
$P_{\rm na}(\tau)$, by considering a point mass and its nearest neighbor as
a two-point-mass lens: $P_{\rm na}(\tau) = 1 - \exp(-8\tau)$.
For $\tau = 0.1$, $P_{\rm na} = 0.55$.
This explains the relatively strong contribution to $P(A)$ by the
collective effect of two (or more) point masses (which results in the
enhanced caustic-induced feature at moderate $A$) for $\tau$ as small as
$0.1$.

To understand the difference between the two- and three-dimensional cases,
we need to estimate $P_{\rm na}(\tau)$ for the three-dimensional case.
We now extend the method used in Paper I to estimate $P_{\rm na}(\tau)$
to the three-dimensional case.
We begin by considering lensing configurations with two point masses on
different lens planes.
Since our results in \S~4.2 and the numerical results of Rauch (1991)
show that $P(A)$ is not sensitive to either the source redshift or
the background cosmology,
we shall simplify the analysis by assuming that $a(\chi_{so}) \approx a_0$.
Then the two point mass configurations of interest are identical to those
studied in \S~3.1 and are specified by two parameters:
the dimensionless distance $\beta_{ij}$ between the lens planes and the
dimensionless Lagrangian separation $d$ between the lenses [the actual
Lagrangian separation is $(x_{E,i}^2 + x_{E,j}^2)^{1/2} d$;
see eq.(\ref{eq7})].
Note that we have changed the notation slightly and the point masses are
on lens planes $i$ and $j$ instead of lens planes $1$ and $2$.
As we saw in \S~3.1, the topology of the caustics (and the critical curves)
changes with decreasing $d$ from type $A$ to $F$.
The only topology with two astroid-shaped primary caustics is type $A$;
the other topology types are produced by the strong effect of both point
masses (see Fig.~6).
In Figure 14 we plot the separation $d_{AB}$ at which the topology changes
from type $A$ to type $B$ as a function of $\beta_{ij}$.
(The data were obtained using the numerical code that generated the
caustics and critical curves shown in Figs.~2--10.)
The separation $d_{AB}$ decreases with increasing $\beta_{ij}$.
In the limit $\beta_{ij} = 0$, the two point masses are on the same plane
and we have $d_{AB} = 2$ (Schneider \& Weiss 1986).
In the limit $\beta_{ij} \to 1$ (i.e., in the limit $i \to s$ or $j \to o$;
see eq.[\ref{eq2}]), $d_{AB} \to 0$;
this is due to the focusing of the light rays by the point mass on lens
plane $i$ and the fact that $x_{E,j}/x_{E,i} \to 0$.

Let us consider first a simple extension of the method used in Paper I to
estimate $P_{\rm na}(\tau)$.
It is convenient to consider the projected distribution (i.e.,
the distribution in Lagrangian space) of the point mass lenses.
If we select randomly one of the point masses, the probability that the
point mass is on lens plane $i$ is $n_i/n$, where $n = \sum_{i=1}^N n_i$ is
the total projected surface number density of point masses.
This point mass has an absolute nearest neighbor in Lagrangian space.
Since the point masses are distributed randomly,
the probability that the absolute nearest neighbor is on lens plane $j$
and at a Lagrangian distance less than $r$ is
$(n_j/n) [1 - \exp(-\pi n r^2)]$.
If we ignore the deflection due the other lenses and consider the
point mass and its nearest neighbor as a two-point-mass lens,
the discussion in the previous paragraph tells us that the primary
caustic associated with the point mass is not an isolated astroid if
$r < (x_{E,i}^2 + x_{E,j}^2)^{1/2} d_{AB}(\beta_{ij})$.
Therefore, an estimate of the fraction of point masses whose caustics are
not isolated astroids is
\begin{equation}
P'_{{\rm na},N} (\tau) = \sum_{i,j=1}^N {n_i n_j \over n^2}
    \left\{1 - \exp\left[-\pi n \left(x_{E,i}^2 + x_{E,j}^2\right)
                         d_{AB}^2(\beta_{ij})\right] \right\} .
\label{eq36}
\end{equation}

In the three-dimensional case, since the (Lagrangian) Einstein radius
$x_{E,L}$ increases from $0$ to $\infty$ as $\chi_{Lo}$ increases (see
eq.[\ref{eq6}]),
it is possible, e.g., for the primary caustic (and critical curve) of a
point mass to merge with that of its second nearest neighbor (from plane
$k$), but not with that of its absolute nearest neighbor (from plane $j$),
if $x_{E,k} \gg x_{E,j}$.
Thus the function $P'_{{\rm na},N} (\tau)$, which takes into account the
absolute nearest neighbor only, probably underestimates the fraction of
point masses whose caustics are not isolated astroids.
We can derive a better estimate by taking into account the nearest
neighbors from all of the lens planes.
For a randomly selected point mass (on lens plane $i$ with probability
$n_i/n$), the probability that the nearest neighbor from lens plane $j$ is
at Lagrangian distance greater than $r$ is $\exp(-\pi n_j r^2)$.
If we ignore the deflection due the other lenses and consider the
point mass and its nearest neighbor from lens plane $j$ as a two-point-mass
lens, the two-point-mass lens produces two astroid-shaped primary caustics
if $r > (x_{E,i}^2 + x_{E,j}^2)^{1/2} d_{AB}(\beta_{ij})$.
Thus the probability that the primary caustic associated with the point
mass is an isolated astroid is $\prod_{j=1}^N \exp[-\pi n_j (x_{E,i}^2 +
x_{E,j}^2) d_{AB}^2(\beta_{ij})]$.
Therefore, a better estimate of the fraction of point masses whose caustics
are not isolated astroids is
\begin{equation}
P_{{\rm na},N} (\tau) = \sum_{i=1}^N {n_i \over n} \left\{
    1 - \exp\biggl[-\pi \sum_{j=1}^N n_j \left(x_{E,i}^2 + x_{E,j}^2\right)
                   d_{AB}^2(\beta_{ij}) \biggr] \right\} .
\label{eq37}
\end{equation}

As with $f_N (A; \tau)$, although $P'_{{\rm na},N}$ and $P_{{\rm na},N}$ are
denoted as functions of $\tau$, they do depend on the location and
the surface density of the lens planes.
For $N = 1$, i.e., for a single lens plane, both $P_{{\rm na},N} (\tau)$
and $P'_{{\rm na},N} (\tau)$ reduce correctly to $P_{\rm na} (\tau) =
1 - \exp(-8\tau)$.
In the limit $\tau \ll 1$,
\begin{equation}
P_{{\rm na},N} (\tau) \approx P'_{{\rm na},N} (\tau)
\approx \tau \sum_{i,j} \left(n_i n_j \over n^2\right)
        {\pi n (x_{E,i}^2 + x_{E,j}^2) \over \tau} d_{AB}^2(\beta_{ij}) .
\end{equation}
In Figure 15 we show $P'_{{\rm na},N} (\tau)$ for $N = 1$, $2$, $4$, $8$,
and $\infty$ (dashed lines) and  $P_{{\rm na},\infty} (\tau)$ (solid line),
using the constant comoving density model adopted in \S~4.2.
Both $P'_{{\rm na},N}$ and $P_{{\rm na},N}$ have converged by $N \approx 16$.
The difference between $P'_{{\rm na},\infty}$ and $P_{{\rm na},\infty}$ is
small for $\tau \la 0.1$, but it increases to $0.10$ at $\tau = 0.5$.

It is clear from Figure 15 that $P_{{\rm na},\infty} (\tau)$ is
significantly smaller than $P_{\rm na} (\tau)$ (the highest dashed curve in
Fig.~15).
In the limit $\tau \ll 1$, $P_{{\rm na},\infty} (\tau) \approx
P'_{{\rm na},\infty} (\tau) \approx 2.5 \tau$ while $P_{\rm na} (\tau)
\approx 8 \tau$.
For $\tau = 0.1$, $P_{{\rm na},\infty} = 0.20$ and $P_{\rm na} = 0.55$;
even for $\tau = 0.233$, $P_{{\rm na},\infty} = 0.40$ only.
Thus, in the three-dimensional case, the fraction of point masses whose
(primary) caustics are not isolated astroids and the contribution to $P(A)$
by configurations of two (or more) point masses are much smaller.
This explains why the low optical depth semi-analytic feature $f_\infty$
provides a reasonably good fit to the caustic-induced feature in $P(A)$ for
$\tau$ as large as $0.233$.
At sufficiently large $\tau$ ($> 0.35$), $P_{{\rm na},\infty} > 0.5$ and
$P(A)$ should deviate significantly from the semi-analytic result.

\subsection{Modification of $P(A)$ at Low Magnification}

Finally, we discuss briefly the deviations of $P(A)$ from the
semi-analytic distribution $K f_\infty/(A^2 - 1)^{3/2}$ at low
magnification (see Fig.~13).
As in the planar case, the superposition of the cross-sections of
individual lenses that results in the caustic-induced feature $f_\infty$ is
only valid for the magnifications seen by the observers close to one of the
astroid-shaped primary caustics.
The semi-analytic distribution also requires modification at low
magnification as it diverges as $(A - 1)^{-3/2}$ at $A = 1$ and is not
normalizable.
In the planar case, an observer not particularly well aligned with any
single lens sees a primary image close to the unperturbed position of the
source and a large number of faint secondary images (one close to each
lens) produced by strongly deflected rays.
In Paper I, we took into account the correlations between the magnifications
of the primary and the secondary images and derived the modification of
$P(A)$ at low magnification.
We have not attempted a similar calculation in the three-dimensional case
because the problem is significantly more complicated.
First, for an observer that is not particularly close to either a primary or
a secondary caustic, there are now $\sim 2^{N_\ast}$ secondary images
(where $N_\ast$ is the number of lenses) because a light ray can reached
the observer after being strongly deflected by one or more lenses.
Although the secondary images that have been strongly deflected by more
than one lens are typically much fainter than those that have been strongly
deflected by only one lens, they are far more numerous and we may have
to account for them.
A more serious problem is the presence of weak caustic-induced features
at low magnification that are associated with the secondary caustics
(see \S~3).
Unlike the features associated with the primary caustics at high
magnification, there are no simple semi-analytic descriptions of the
features associated with the secondary caustics.

\section{SUMMARY}

In this paper and Paper I, we have sought to develop a systematic theory
of gravitational microlensing.
We focused on microlensing by a planar distribution of point masses in
Paper I.
In this paper we have studied gravitational microlensing by a
three-dimensional distribution of point masses, focusing on the nature of
the critical curves, the caustics, the illumination pattern, and especially
the macroimage magnification distribution $P(A)$.
In the process, we have explained many of the differences between two- and
three-dimensional microlensing that have been seen in previous numerical
simulations.

In \S~3 we studied the lensing properties of three-dimensional lens
distributions by considering in detail multiplane configurations with two,
three, and four point masses.
The lensing properties of a three-dimensional lens distribution is
fundamentally different from that of a planar distribution in that a light
ray could have arrived in the vicinity of a lens either having suffered
only weak deflections along the way from the source, or after having been
strongly deflected one or more times by the intervening lens.
The latter situation cannot arise in a purely planar distribution of lenses,
and it gives rise to secondary caustics and critical lines that have no
analogues in the planar case.
In addition, in the case of more than two point masses, the caustic
topology is no longer simply related to the critical curve topology because
caustics corresponding to different critical lines may overlap each other.

For $N_\ast$ point masses that are widely separated in Lagrangian space,
there are $\sim N_\ast$ primary critical curves (i.e., critical curves that
have not suffered any strong deflections) and $\sim 2^{N_\ast} - 1$ critical
curves in total.
Features in the ``normalized'' differential cross-section, $\varphi(A)$,
that are associated with the secondary caustics are relatively weak and
at low magnifications.
Significant features occuring at moderate to high magnifications are
invariably associated with the primary caustics.

In \S~4 we derived a semi-analytic expression for the caustic-induced
feature at the high magnification end of $P(A)$ for a low optical depth
lens distribution.
In the low optical depth limit, we showed that the multiplane lens equation
near a point mass is formally equivalent to the single plane equation of a
point mass perturbed by shear.
Consequently, the (primary) caustic has the familiar astroid shape, and
the differential cross-section exhibits the same scaling behavior found in
Paper I (i.e., $\varphi$ depends on magnification $A$ and the effective
shear $S'$ through the combination $AS'$).
By modeling the illumination pattern as a superposition of the patterns
due to individual ``point mass plus weak shear'' lenses,
we found that the caustic-induced feature in the macroimage magnification
distribution $P(A)$ can be written as a weighted sum of the semi-analytic
feature derived in Paper I for a planar lens distribution (eq.[\ref{eq32}]).
The resulting semi-analytic caustic-induced feature is similar to the
feature in the planar case, but it is shifted to higher magnification.

In the moderate to high $A$ regime, our semi-analytic $P(A)$ agrees
remarkably well with the magnification distribution obtained from numerical
simulations by Rauch (1991), even for optical depth as high as
$\tau \sim 0.2$.
This is very different from the two-dimensional case studied in Paper I.
In the two-dimensional case, we have found that the caustic-induced feature
in $P(A)$ tends to be more pronounced than the low optical depth result,
even for optical depth as low as $\tau \sim 0.1$.
We argued that this particular difference between two- and three-dimensional
microlensing is due to the fact that for a given value of optical depth,
the fraction of point masses whose (primary) caustics are not simple
astroids is significantly higher in the planar case than in the
three-dimensional case.
Finally, we discussed briefly the deviations of the numerical $P(A)$ from
the semi-analytic distribution at low magnification.

\acknowledgments
We thank K.~Rauch for providing us with the results of Rauch (1991).
We also thank S.~Mao and K.~Rauch for helpful discussions.
A.B.\ and L.K.\ acknowledge the hospitality shown to them at CITA during
their visits in 1995.
This work was supported in part by NSERC (Canada), the CIAR cosmology
program, CITA, the Institute for Astronomy (L.K.), the Dudley Observatory
(A.B.), and a CITA National Fellowship (M.H.L.).

\clearpage
\appendix

\section{STATISTICS OF SHEAR FOR A LOW OPTICAL DEPTH\\
         THREE-DIMENSIONAL LENS DISTRIBUTION}

In this Appendix we evaluate $p(S'|\chi_{Lo})$, the probability distribution
of the effective shear $S'(\chi_{Lo})$ at a given $\chi_{Lo}$, due to
a low optical depth three-dimensional distribution of point masses.
It is assumed that the point masses are distributed randomly on $N$ lens
planes, each with projected (onto the observer plane) surface number density
$n_i$.
Specifically, on each lens plane, $N_i = \pi n_i R_i^2$ point masses are
distributed randomly within a circle of Lagrangian radius $R_i$.

At a given $\chi_{Lo}$, the effective shear (in matrix form) is
(eq.[\ref{eq20b}])
\begin{equation}
\shear'_L = S'(\chi_{Lo}) \tshear(\phi'_L)
          = \sum_{i=1}^L \left(\chi_{si}\chi_{Lo}\over
                               \chi_{sL}\chi_{io}\right) \shear_i +
            \sum_{i=L+1}^N \left(\chi_{sL}\chi_{io}\over
                                 \chi_{si}\chi_{Lo}\right) \shear_i ,
\end{equation}
where (eq.[\ref{eq12b}])
\begin{equation}
\shear_i = x^2_{E,i} \sum_{k=1}^{N_i} {1 \over |{\vct x}_{ik}|^2}
           \left(\begin{array}{lr}
           \cos 2\phi_{ik} & \sin 2\phi_{ik}\\
           \sin 2\phi_{ik} &-\cos 2\phi_{ik}
           \end{array}\right).
\end{equation}
It is convenient to represent the two distinct components of the shear matrix
$\shear_i$ as a vector:
\begin{equation}
{\vct S}_i = x^2_{E,i} \sum_{k} {\vct s}_{ik},
\end{equation}
where ${\vct s}_{ik} =
(\cos2\phi_{ik}/|{\vct x}_{ik}|^2,\, \sin2\phi_{ik}/|{\vct x}_{ik}|^2)$.
Then the effective shear $\shear'_L$ can be written as an appropriately
weighted vector superposition:
\begin{eqnarray}
{\vct S}'(\chi_{Lo})
&=& \sum_{i=1}^{L} \left(\chi_{si}\chi_{Lo}\over\chi_{sL}\chi_{io}\right)
    x^2_{E,i} \sum_{k} {\vct s}_{ik} +
    \sum_{i=L+1}^N \left(\chi_{sL}\chi_{io}\over\chi_{si}\chi_{Lo}\right)
    x^2_{E,i} \sum_{k} {\vct s}_{ik} \nonumber \\
& & \\
&\equiv& \sum_{i=1}^{N} G_i \sum_{k} {\vct s}_{ik} ,\nonumber
\end{eqnarray}
where we have defined $G_i$ to simplify the notations.

Since all the ${\vct s}_{ik}$ are statistically independent and have similar
statistical properties, the formal expression
for the probability distribution of ${\vct S}'(\chi_{Lo})$ is
\begin{equation}
p({\vct S}'|\chi_{Lo}) = \int\!\!\cdots\!\!\int \prod_{i=1}^N \prod_{k} \,
d^2{\vct s}_{ik} \, p({\vct s}_{ik})\,
\delta^2\left({\vct S}' - \sum_{i=1}^{N} G_i \sum_{k} {\vct s}_{ik}\right),
\end{equation}
where $\delta^2({\vct s})$ is the two-dimensional delta function.
To evaluate equation (A5), it is convenient to express the delta
function and $p({\vct s}_{ik})$ as Fourier transforms:
\begin{equation}
\delta^2 \left({\vct S}' - \sum_{i=1}^{N} G_i \sum_{k} {\vct s}_{ik}\right)
= {1\over (2\pi)^2} \int d^2{\vct u} \,
e^{-i {\vct u} \cdot \left({\vct S}' - \sum_{i=1}^{N} G_i \sum_{k}
                           {\vct s}_{ik}\right)};
\end{equation}
\begin{equation}
p({\vct s}_{ik}) = {1\over (2\pi)^2} \int d^2{\vct t}_{ik} \, q({\vct t}_{ik})
\, e^{-i{\vct t}_{ik}\cdot{\vct s}_{ik}},
\end{equation}
where the characteristic function $q({\vct t}_{ik})$ is 
(Nityananda \& Ostriker 1984)
\begin{equation}
q({\vct t}_{ik}) = 1 - {\pi n_i\over N_i}t_{ik}.
\end{equation}
Substituting equations (A6) and (A7) into equation (A5), we can perform the
integration with respect to ${\vct s}_{ik}$:
\begin{equation}
{1\over (2\pi)^2} \int d^2{\vct s}_{ik} \, e^{-i{\vct s}_{ik}\cdot\left(
{\vct t}_{ik}-G_i{\vct u}\right)}
= \delta^2 \left({\vct t}_{ik}-G_i{\vct u}\right).
\end{equation}
Next, we integrate with respect to ${\vct t}_{ik}$, grouping together the 
$N_i$ integrals associated with the $i$th screen:
\begin{equation}
\int\!\!\cdots\!\!\int \prod_{k=1}^{N_i} \, d^2{\vct t}_{ik} \,
q({\vct t}_{ik}) \, \delta^2\left({\vct t}_{ik}-G_i{\vct u}\right)
= \left(1 - {\pi n_i G_i u\over N_i}\right)^{N_i}
\longrightarrow e^{-\pi n_i G_i u}
\end{equation}
in the limit $N_i \to \infty$.
Hence,
\begin{equation}
p({\vct S}'|\chi_{Lo}) = {1\over (2\pi)^2} \int d^2{\vct u} \,
e^{-\tau' u} e^{-i{\vct u}\cdot{\vct S}'},
\end{equation}
where the ``effective'' optical depth $\tau'(\chi_{Lo})$ is
\begin{equation}
\tau'(\chi_{Lo}) = \pi \sum_i G_i n_i = \sum_{i=1}^{L}
\left(\chi_{si}\chi_{Lo}\over\chi_{sL}\chi_{io}\right) \pi n_i x^2_{E,i} +
\sum_{i=L+1}^N \left(\chi_{sL}\chi_{io}\over\chi_{si}\chi_{Lo}\right)
\pi n_i x^2_{E,i} .
\end{equation}
Performing the integration with respect to ${\vct u}$ in equation (A11),
we obtain
\begin{eqnarray}
p({\vct S}'|\chi_{Lo})
&=& {1\over 2\pi} \int_0^\infty du \, u \, e^{-\tau' u} J_0(S'u) \nonumber \\
& & \\
&=& {1\over 2\pi} {\tau' \over \left(\tau'^2 + S'^2\right)^{3/2}}. \nonumber
\end{eqnarray}
Since $p({\vct S}'|\chi_{Lo})$ depends only on the magnitude of the
effective shear $S'(\chi_{Lo})$, it is convenient to consider the
probability distribution for the latter:
\begin{equation}
p(S'|\chi_{Lo}) =
{\tau'(\chi_{Lo}) S' \over \left[\tau'^2(\chi_{Lo}) + S'^2\right]^{3/2}}.
\end{equation}
We have expressed the probability distribution explicitly as a conditional
probability that depends on the value of $\chi_{Lo}$.
It should be noted that $p(S'|\chi_{Lo})$ has the same functional form as
the shear distribution due to a planar random distribution of point masses
(eq.[30] of Paper I).

\clearpage
 
\begin{deluxetable}{cccc}
\tablecaption{Configurations with Two Point Masses}
\tablewidth{0pt}
\tablehead{
\colhead{$\beta_{12}$} & \colhead{$d$} & \colhead{Topology} &
\colhead{Pixel Sizes}
}
\startdata
0.0 & 2.4 & $A'$         & 0.020, 0.032\nl
0.0 & 1.0 & $B'$         & 0.016, 0.024\nl
0.0 & 0.6 & $C'$         & 0.016, 0.024\nl
0.3 & 2.0 & $A$\phm{$'$} & 0.020, 0.040\nl
0.3 & 1.0 & $B$\phm{$'$} & 0.018, 0.036\nl
0.3 & 0.6 & $C$\phm{$'$} & 0.016, 0.032\nl
0.3 & 0.5 & $D$\phm{$'$} & 0.016, 0.032\nl
0.3 & 0.4 & $E$\phm{$'$} & 0.016, 0.032\nl
0.3 & 0.2 & $E$\phm{$'$} & 0.020, 0.040\nl
\enddata
\end{deluxetable}

\clearpage

\begin{deluxetable}{ccclrc}
\tablecaption{Configurations with Three and Four Point Masses}
\tablewidth{0pt}
\tablehead{
\colhead{Case} & \colhead{$N_\ast$} & \colhead{$i$} &
\colhead{$\chi_{si}/\chi_{so}$} & \colhead{${\vct x}_{i1}$} &
\colhead{Pixel Sizes}
}
\startdata
a & 3 & 1 & 0.1 & $(-0.50,0.00)$ & 0.010, 0.024\nl
  &   & 2 & 0.5 & $(0.20,0.00)$  & \nl
  &   & 3 & 0.9 & $(0.30,0.25)$  & \nl
\tablevspace{8pt}
b & 3 & 1 & 0.55032 & $(0.35,0.00)$   & 0.012, 0.024\nl
  &   & 2 & 0.79370 & $(-0.40,0.65)$  & \nl
  &   & 3 & 0.94104 & $(-0.30,-0.50)$ & \nl
\tablevspace{8pt}
c & 4 & 1 & 0.5     & $(0.30,0.00)$  & 0.012, 0.024\nl
  &   & 2 & 0.72112 & $(0.00,0.65)$  & \nl
  &   & 3 & 0.85499 & $(-0.80,0.00)$ & \nl
  &   & 4 & 0.95647 & $(0.00,-0.50)$ & \nl
\enddata
\end{deluxetable}

\clearpage

\figcaption[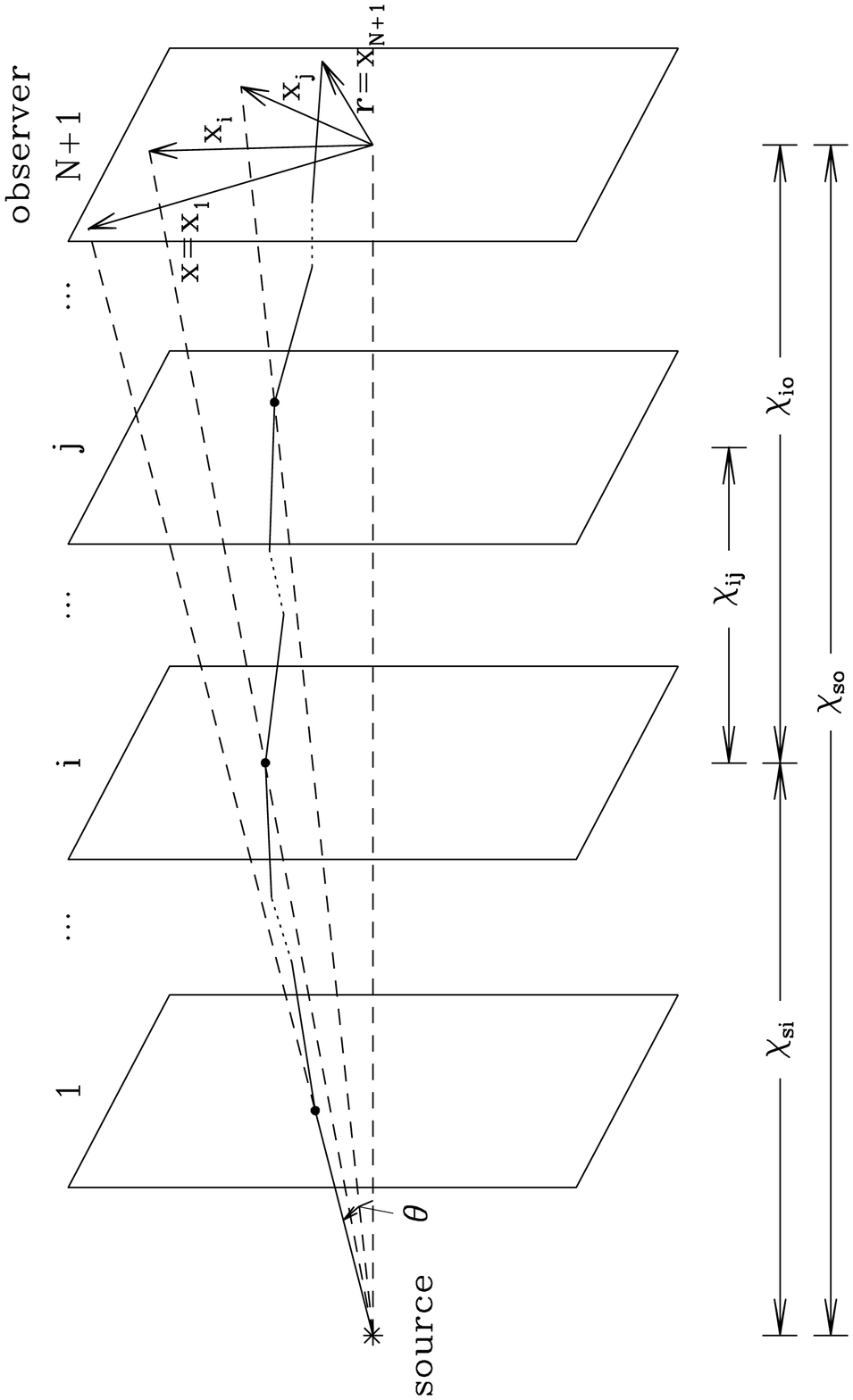]{A schematic diagram illustrating our notation for
multiplane gravitational lensing.
A light ray that leaves the source with angle $\vct\theta$ has Lagrangian
coordinates $\vct x$($= {\vct x}_1$).
It is deflected on the $N$ lens planes and reaches the observer plane
at Eulerian coordinates $\vct r$($= {\vct x}_{N+1}$).}

\figcaption[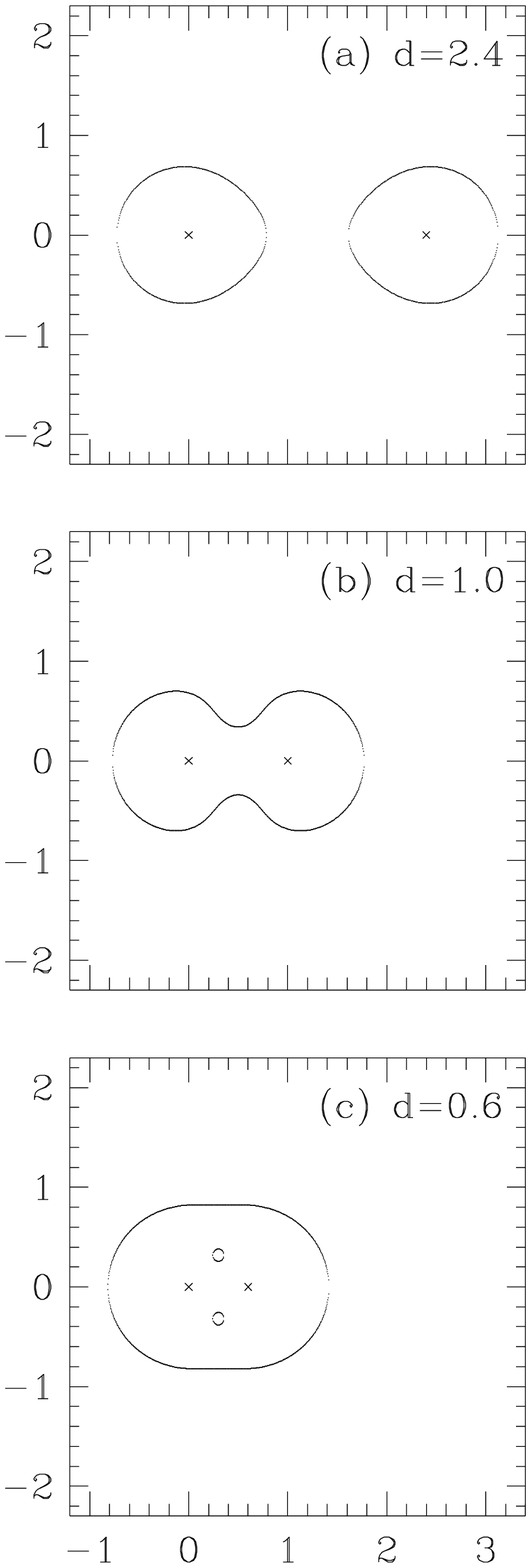]{The critical curves of the configurations with two point
masses ($m_1' = 1 - m_2' = 1/[2-\beta_{12}]$).
The dimensionless distance between the lens planes $\beta_{12} = 0$
(i.e., the two point masses are on the same plane).
Each panel is labeled by the Lagrangian separation, $d$, between the two
point masses, whose Lagrangian positions are indicated by the crosses.}

\figcaption[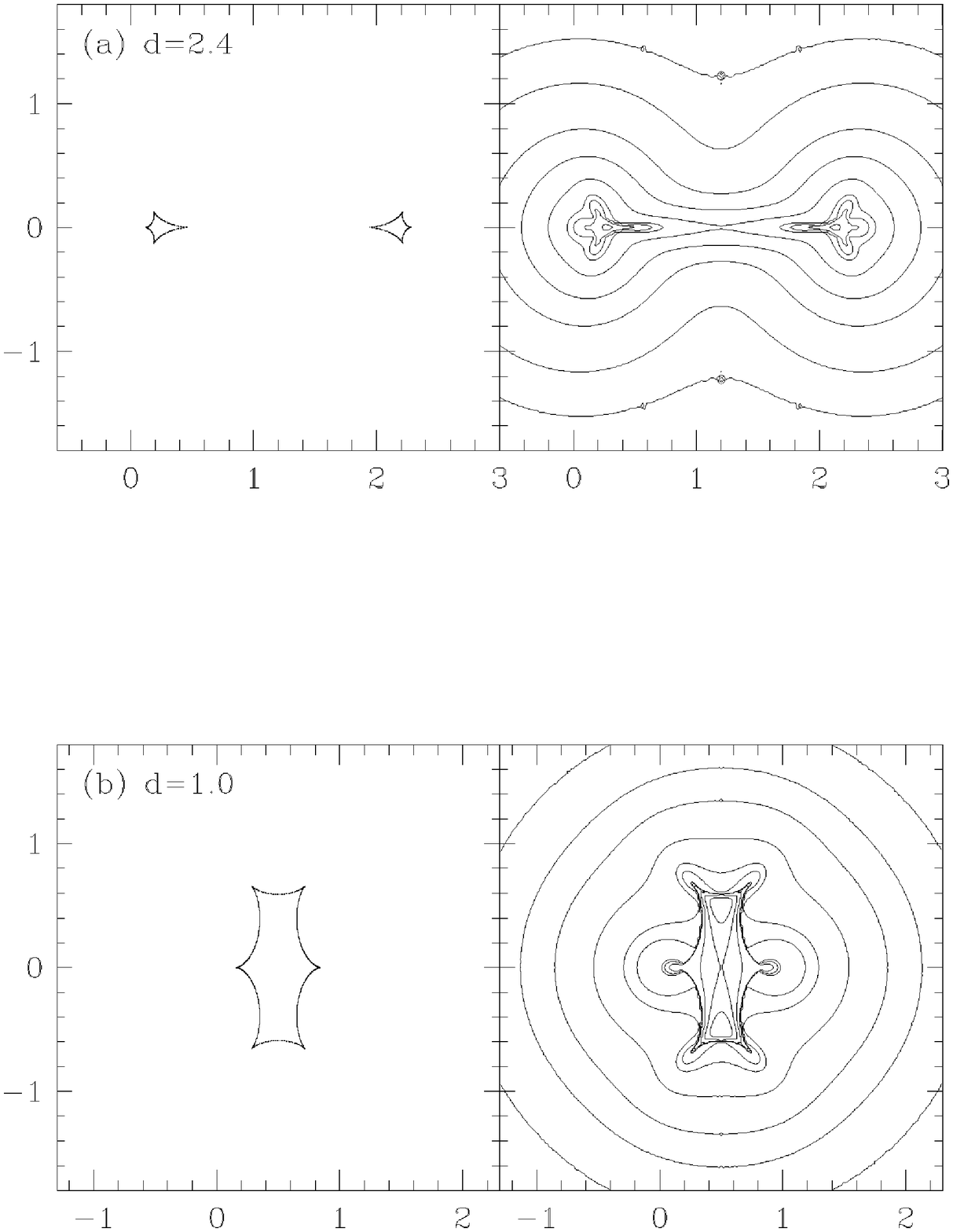, 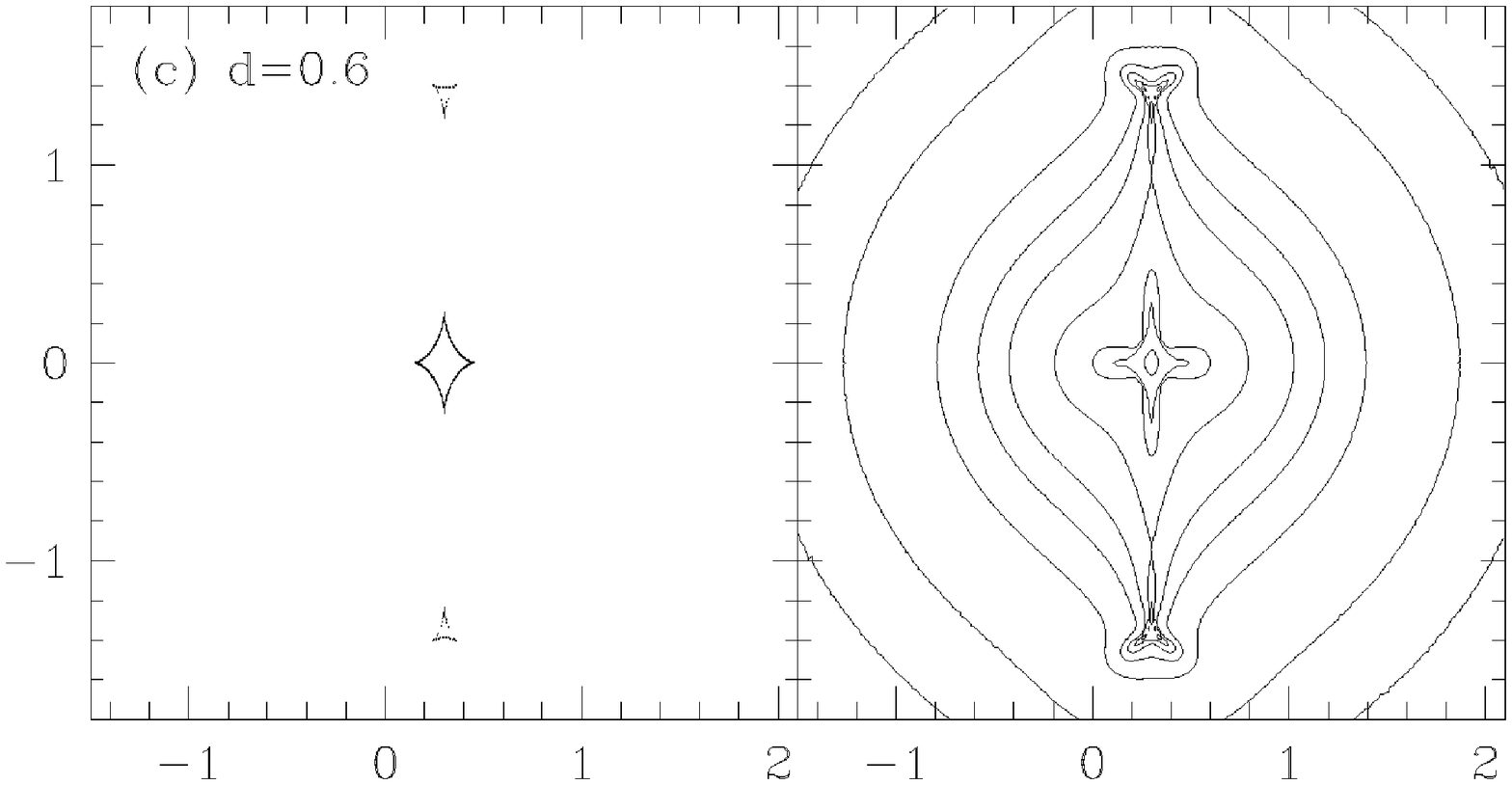]{The caustics ({\it left panels}) and the
iso-magnification
contours ({\it right panels}) of the configurations shown in Fig.~2.
The contour levels are chosen to highlight interesting features in the
magnification patterns.}

\figcaption[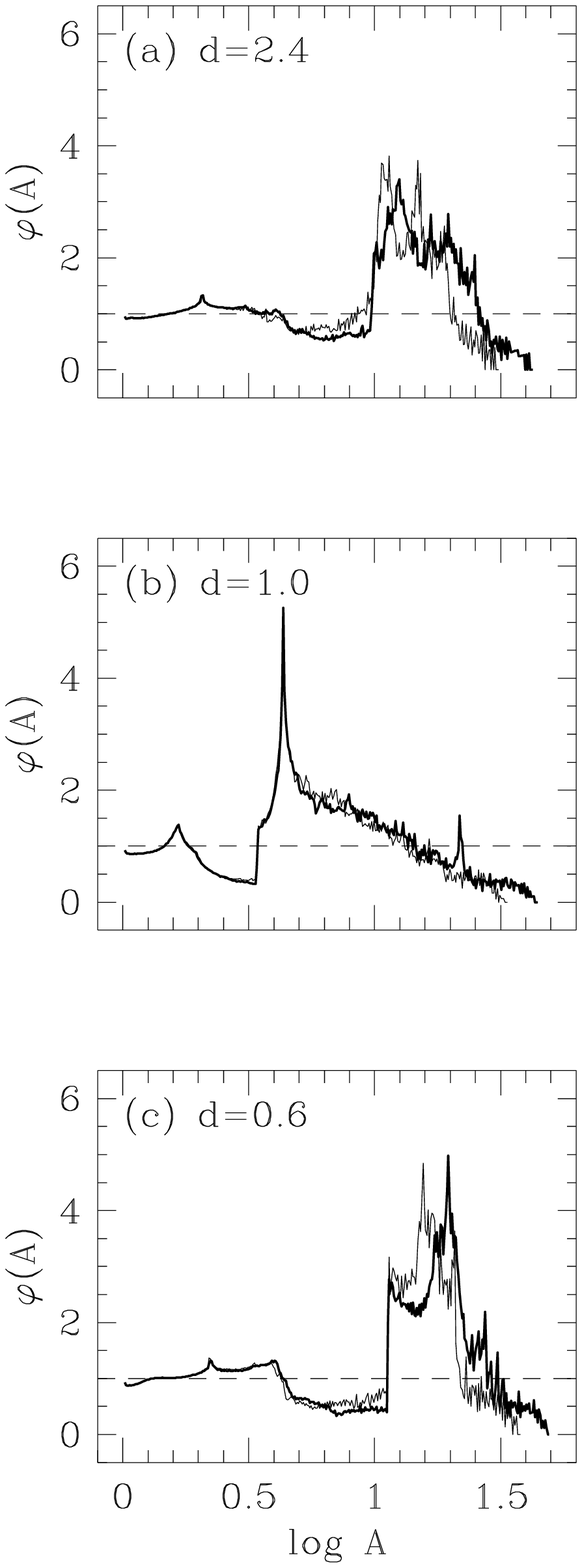]{The ``normalized'' differential cross-sections,
$\varphi(A)$, of the configurations shown in Fig.~3.
Note that $\varphi(A) = 1$ for a single point mass of unit mass.
In each panel, the thick (thin) line was obtained from the ray-shooting
calculation with the smaller (larger) pixel size listed in Table 1.}

\figcaption[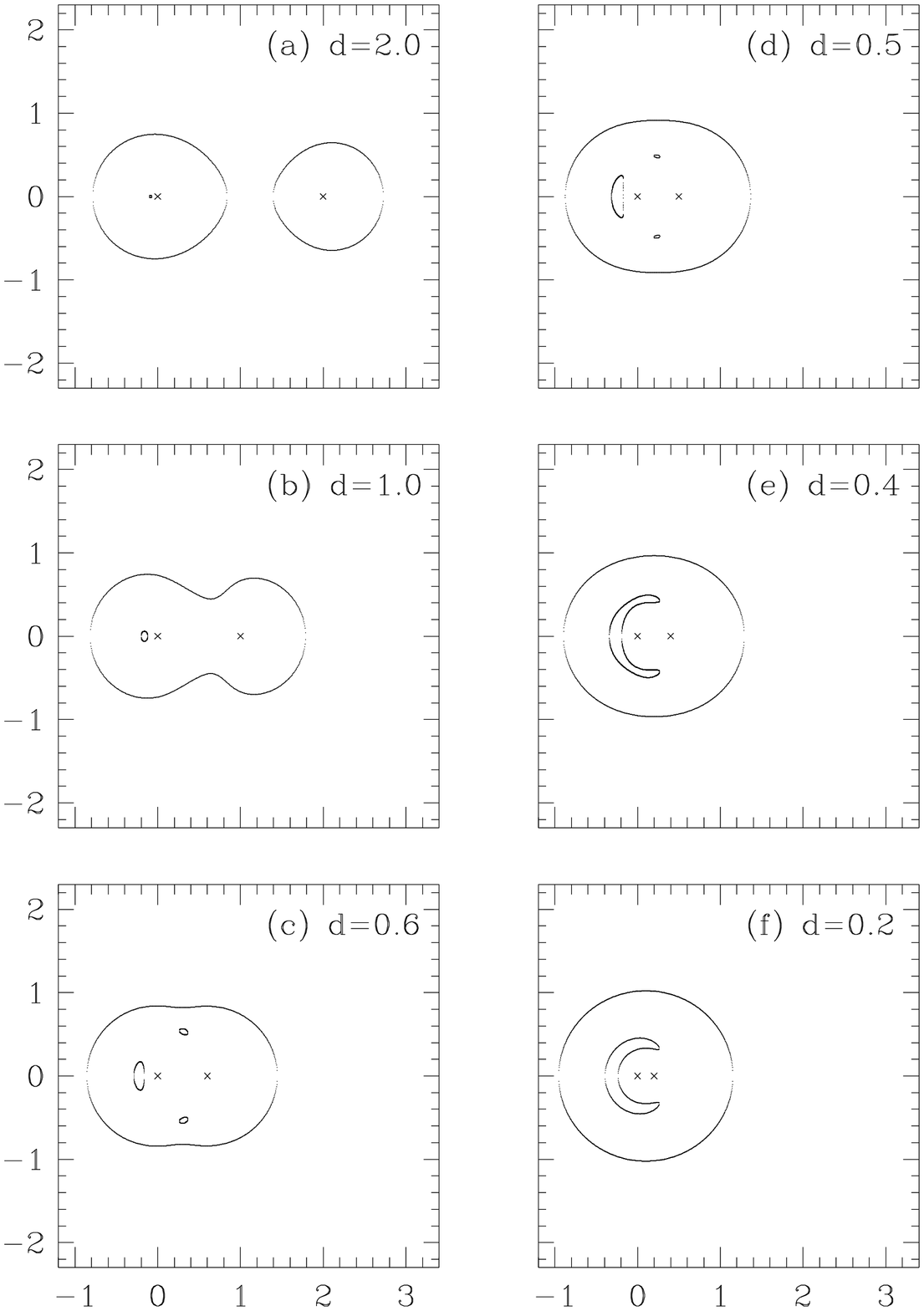]{Same as Fig.~2, but for the configurations with
$\beta_{12} = 0.3$.}

\figcaption[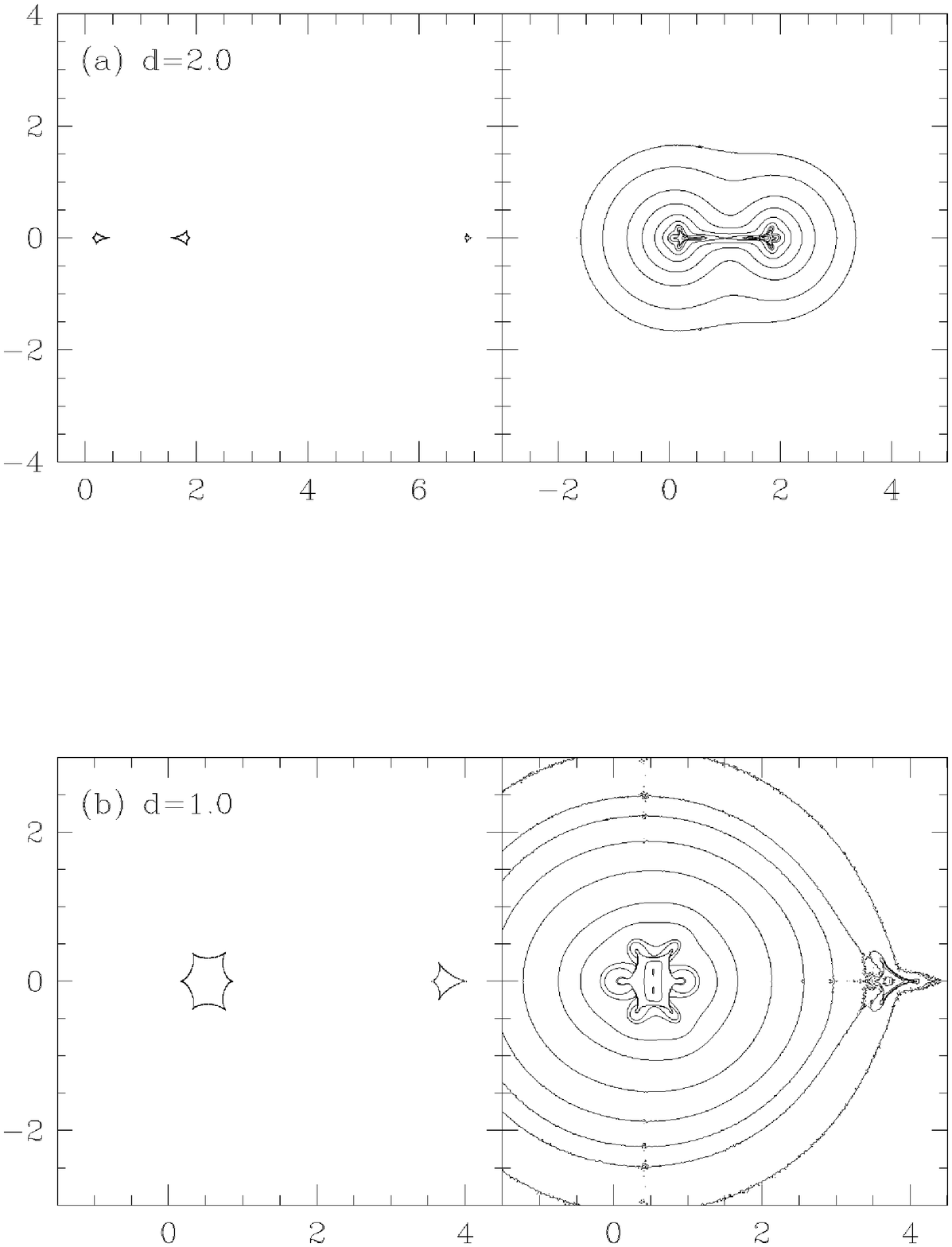, 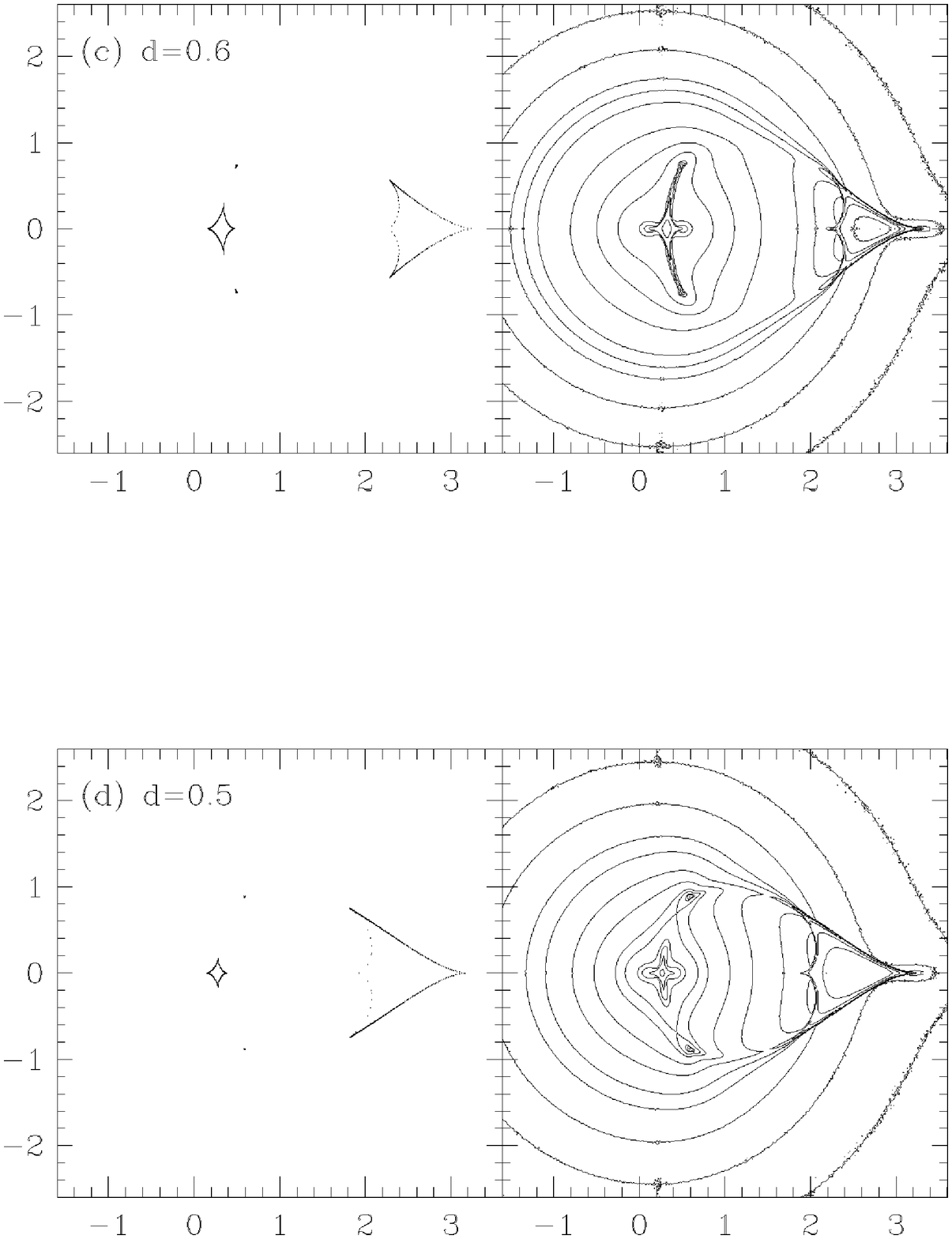, 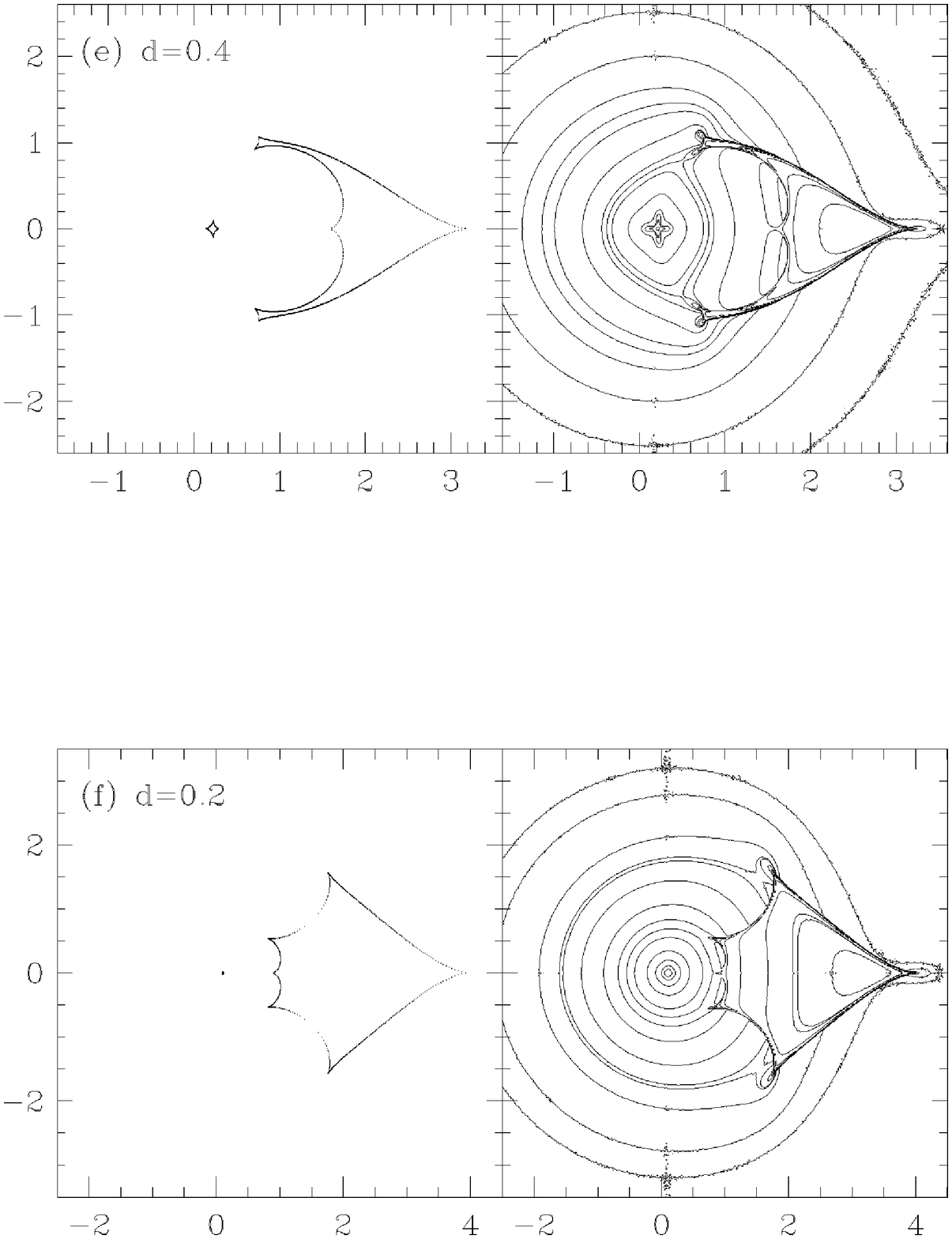]{Same as Fig.~3, but for the
configurations shown in Fig.~5.
Note that there is an offset between the left and right panels of ({\it a}).}

\figcaption[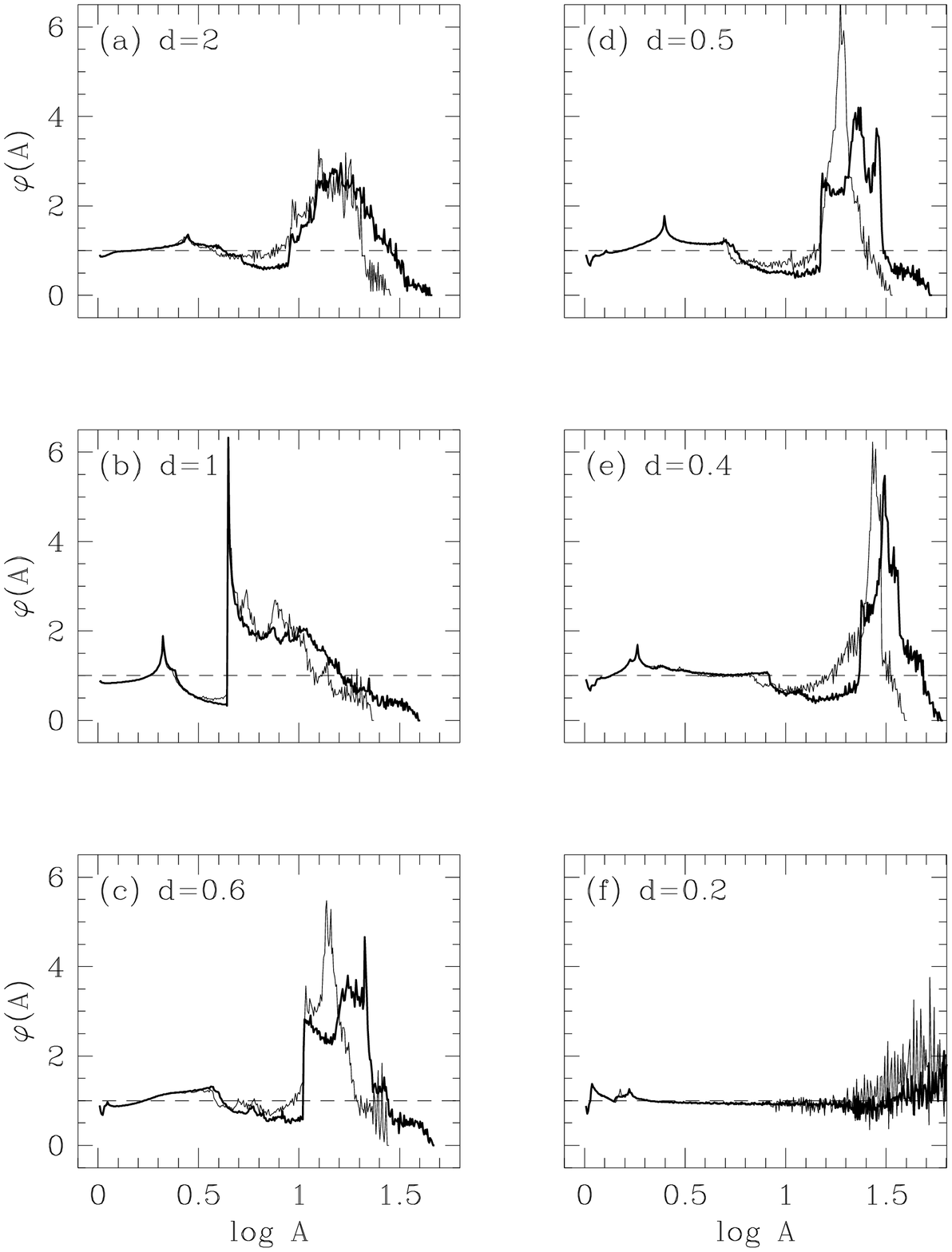]{Same as Fig.~4, but for the configurations shown in
Fig.~6.}

\figcaption[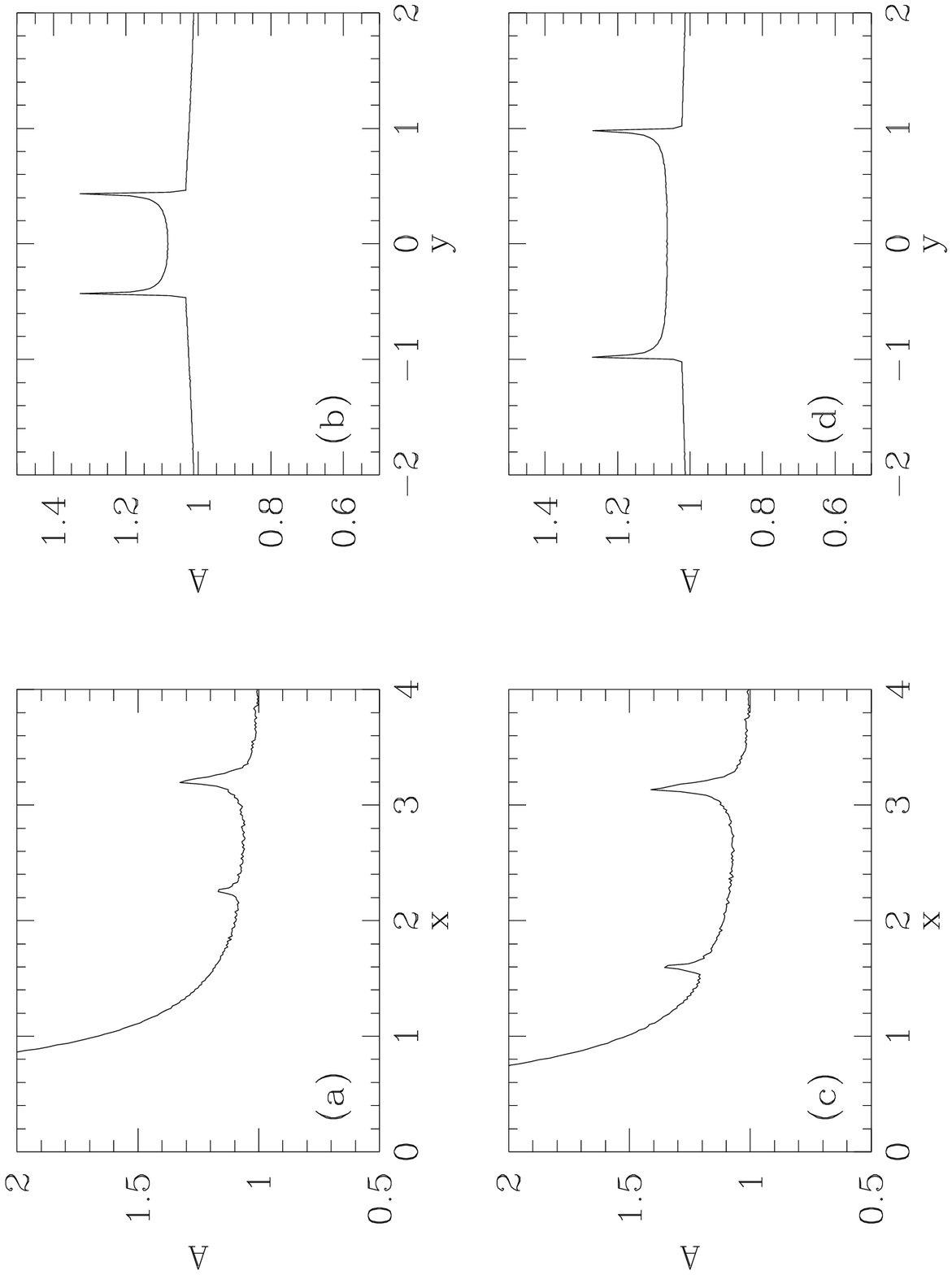]{({\it a}) The magnification factor $A$ as a function of
the $x$-coordinate for the magnification pattern shown in the right panel
of Fig.~6{\it c} and $y = 0$.
({\it b}) The magnification factor $A$ as a function of the $y$-coordinate
for the magnification pattern shown in the right panel of Fig.~6{\it d}
and $x = 2.3$.
({\it c}) Same as ({\it a}), but for Fig.~6{\it e}.
({\it d}) Same as ({\it b}), but for Fig.~6{\it f} and $x = 2.4$.}

\figcaption[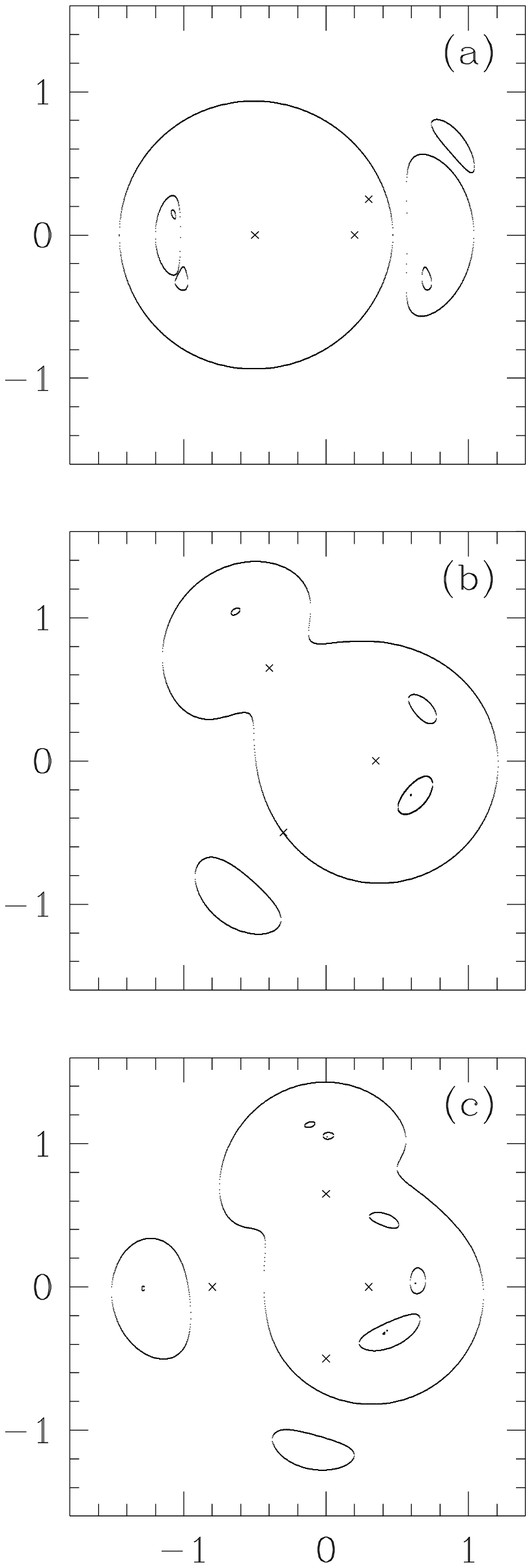]{The critical curves of the configurations ({\it a}) a,
({\it b}) b, and ({\it c}) c of Table 2.
The Lagrangian positions of the point masses are indicated by the crosses.}

\figcaption[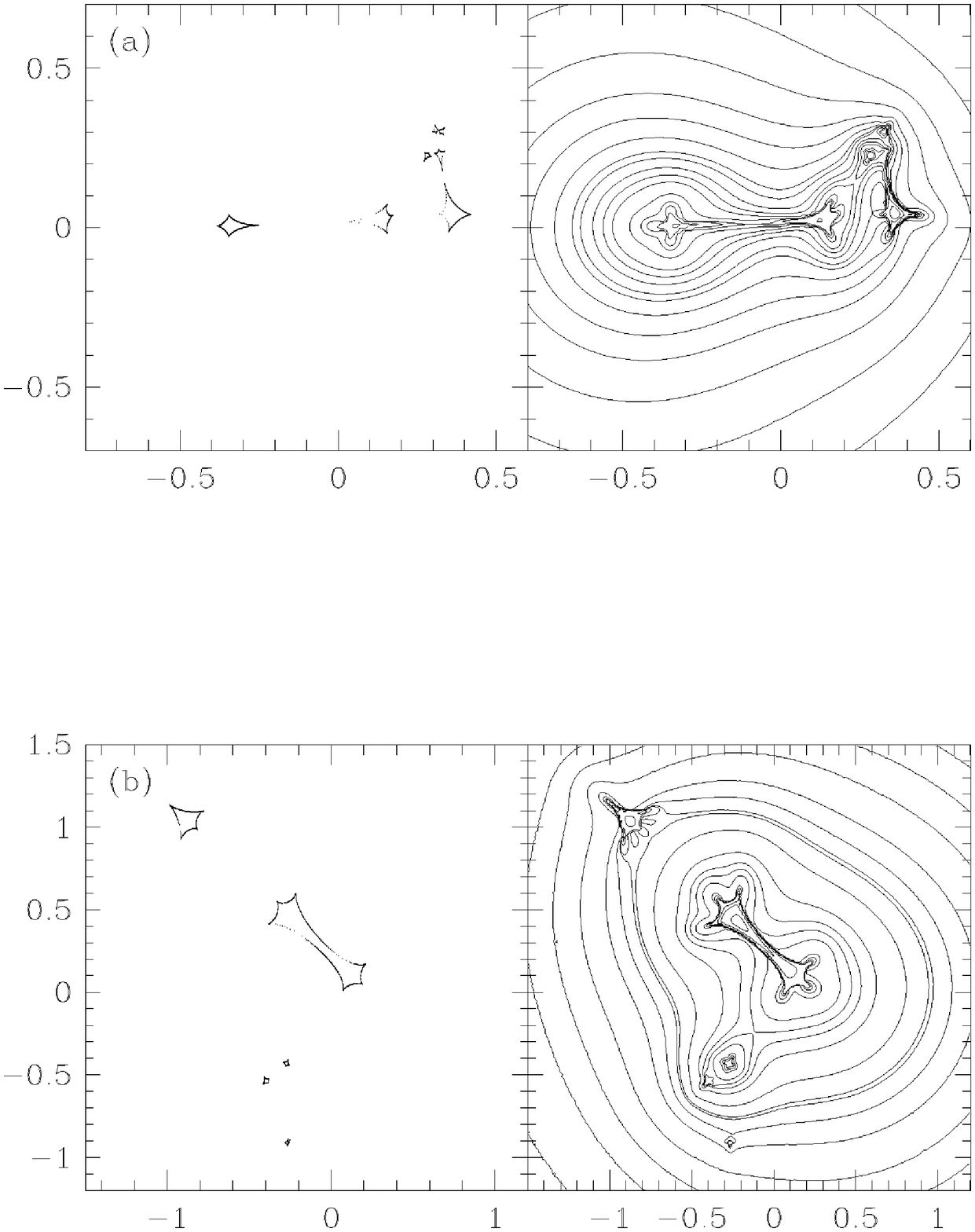, 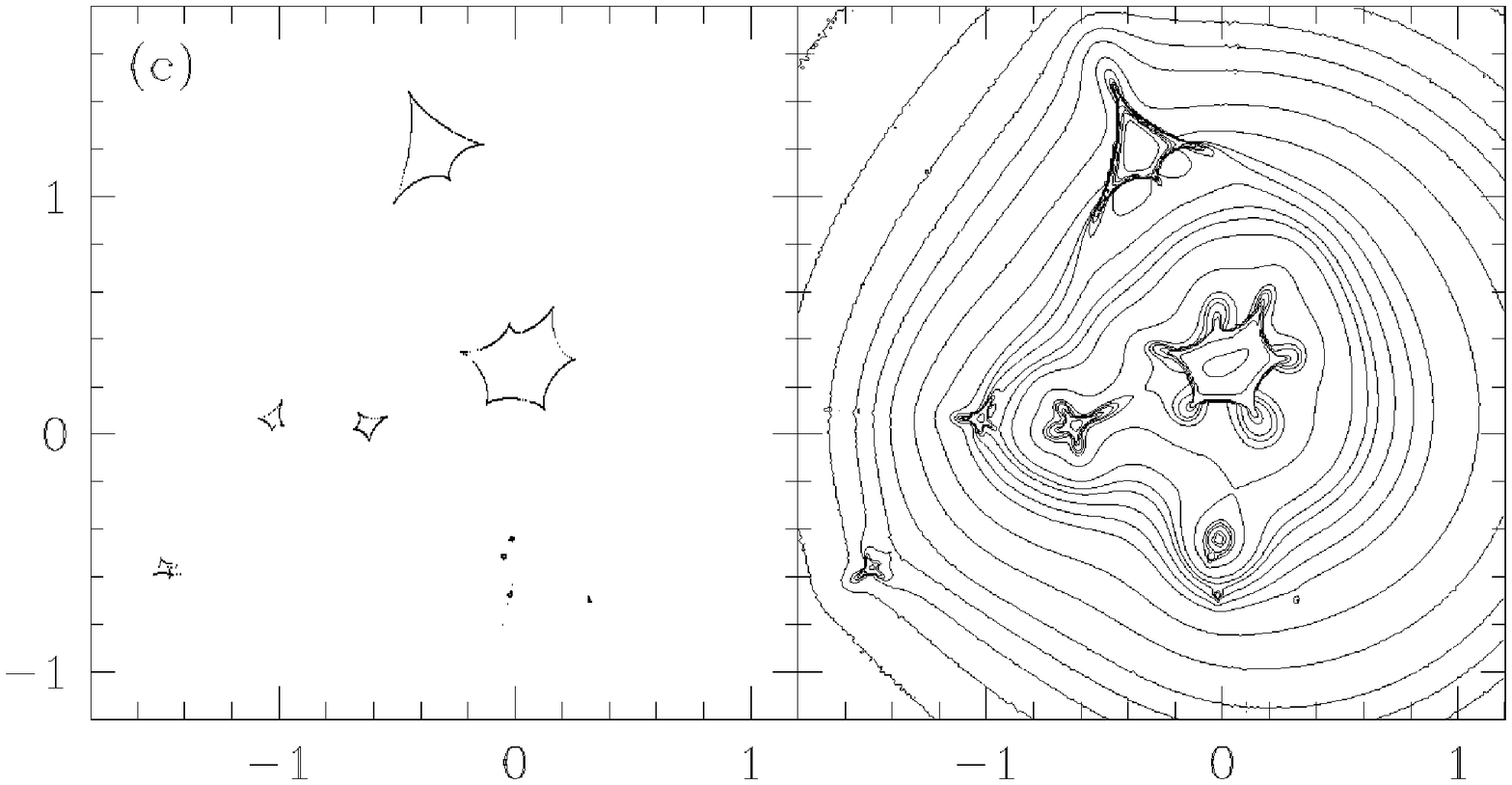]{The caustics ({\it left panels}) and the
iso-magnification contours ({\it right panels}) of the configurations shown
in Fig.~9.
The contour levels are chosen to highlight interesting features in the
magnification patterns.}

\figcaption[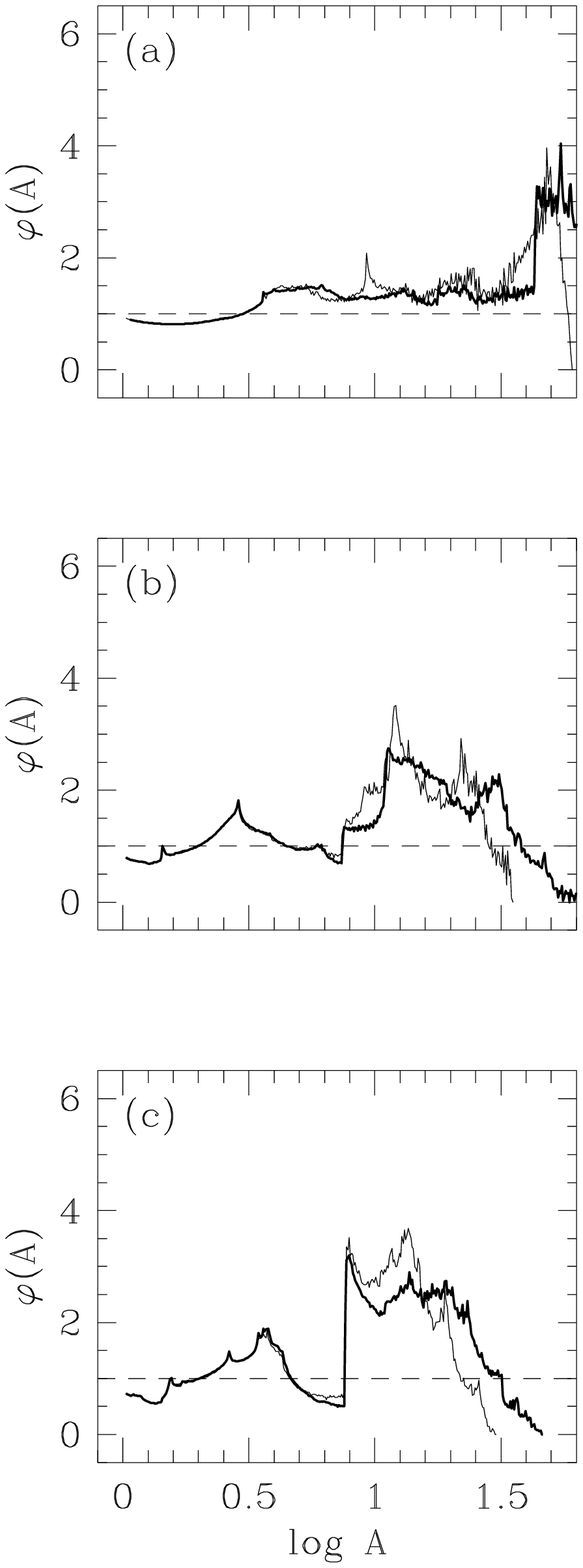]{The ``normalized'' differential cross-sections,
$\varphi(A)$, of the configurations shown in Fig.~10.
In each panel, the thick (thin) line was obtained from the ray-shooting
calculation with the smaller (larger) pixel size listed in Table 2.}

\figcaption[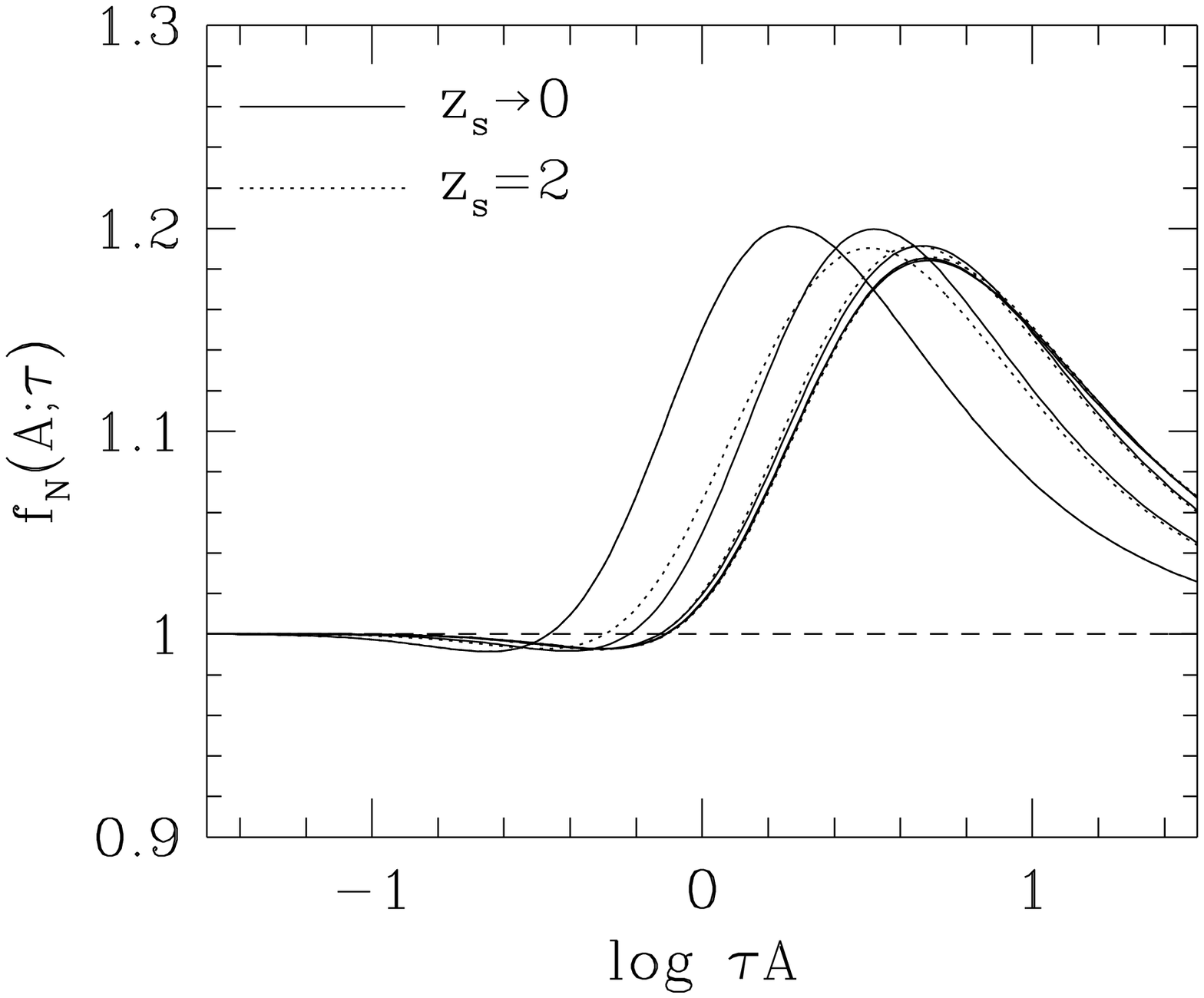]{The function $f_N (A; \tau)$ (eq.[\ref{eq32}]) that
describes the caustic-induced feature in the macroimage magnification
distribution, $P(A)$, at low optical depth $\tau$.
The solid (dashed) lines from left to right show $f_N$ for $N = 1$, $2$,
$4$, $8$, and $16$, and for source redshift $z_s \to 0$ ($z_s = 2$).
A lens distribution with constant comoving mass density is assumed
(see \S~4.2).}

\figcaption[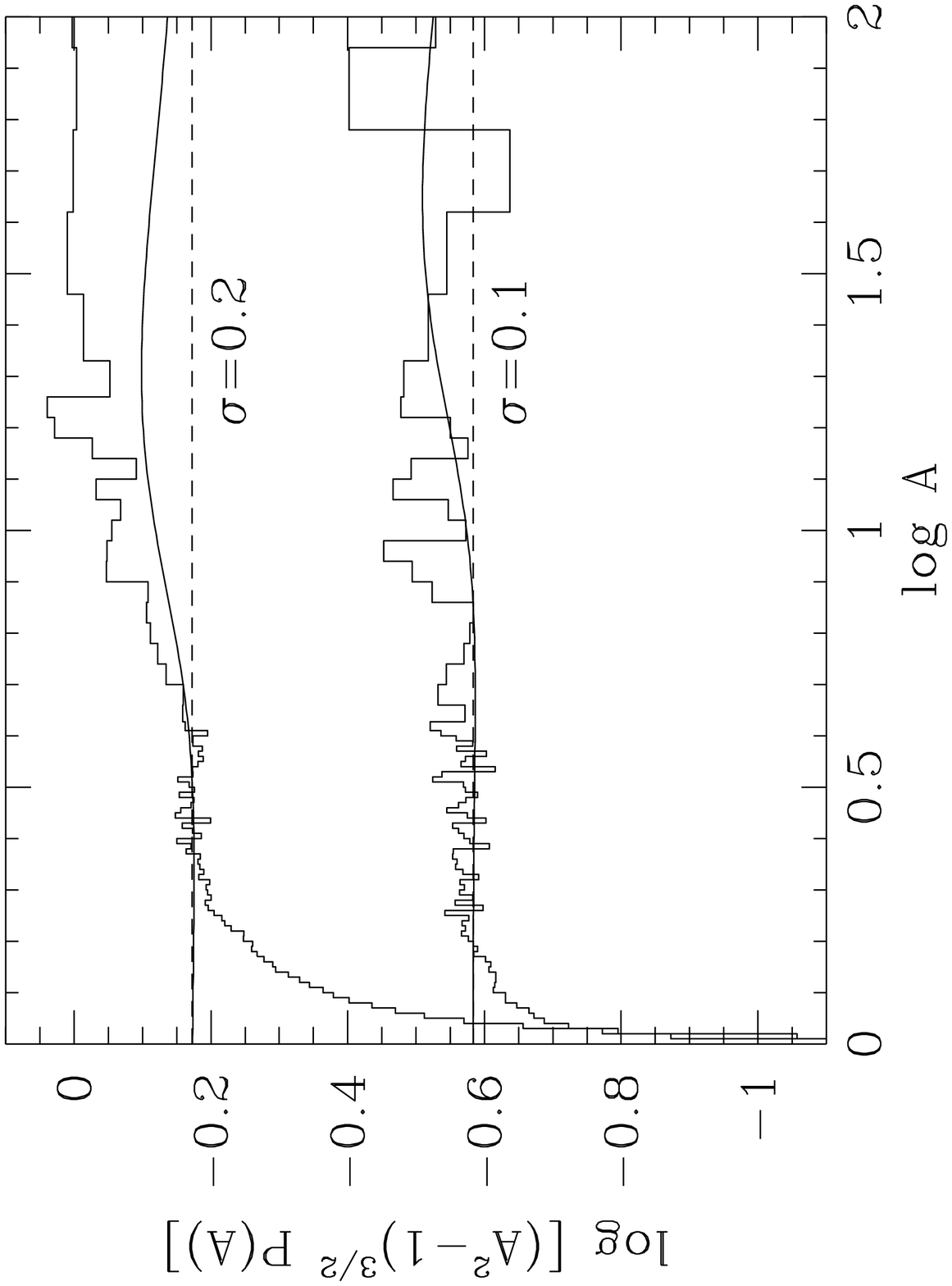]{Comparison of the semi-analytic macroimage magnification
distribution $P(A) = K f_\infty(A; \tau)/(A^2 - 1)^{3/2}$ ({\it solid lines})
with the Monte Carlo results of Rauch (1991; {\it histograms}) for
$\sigma = 0.1$ and $0.2$ (or optical depth $\tau = 0.108$ and $0.233$).
We adopt equation (\ref{eq35}) for the normalization $K$ ({\it dashed lines}).}

\figcaption[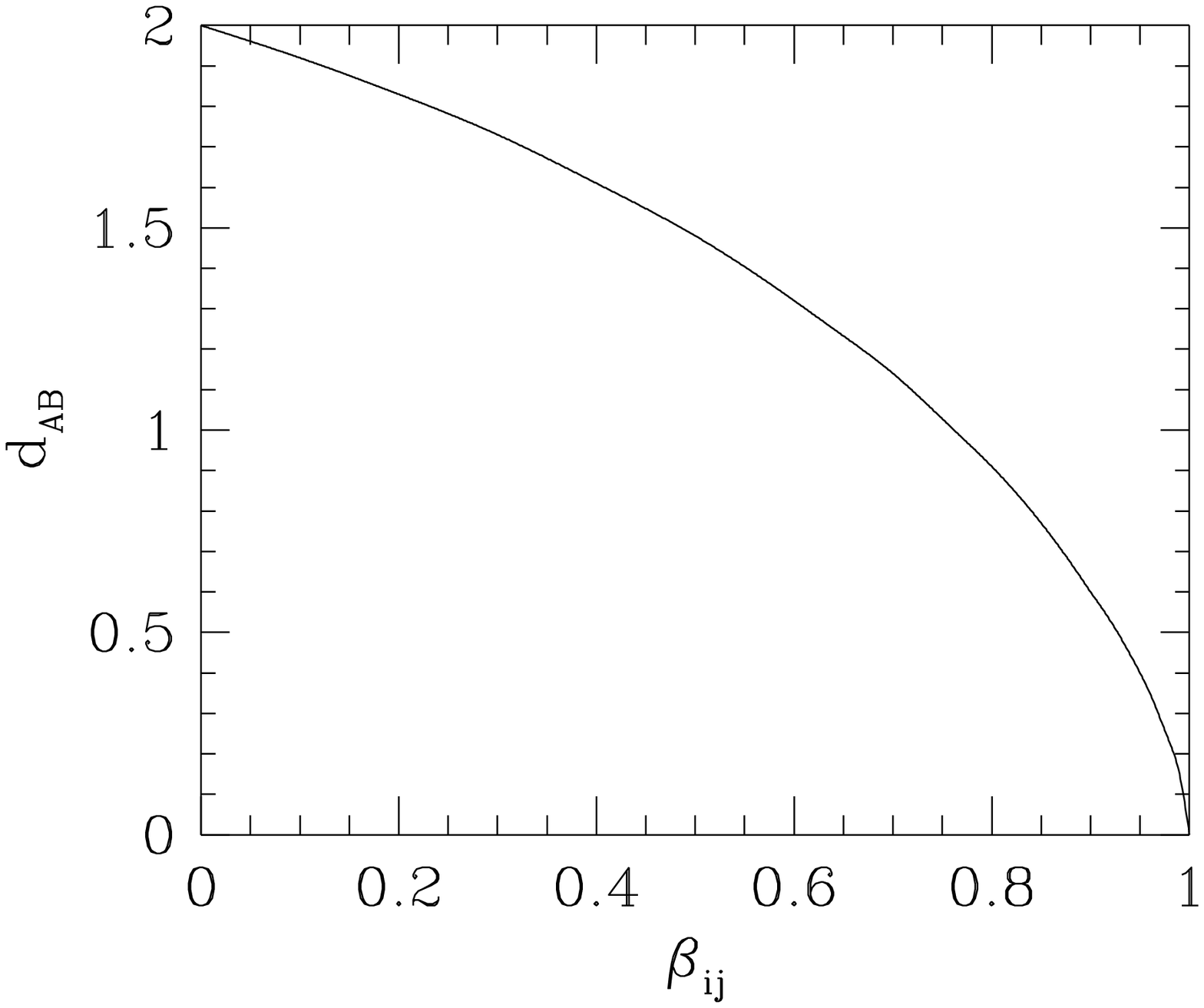]{The Lagrangian separation $d_{AB}$ at which the critical
curve (and caustic) topology of the two-point-mass lens changes from type
$A$ to type $B$ as a function of $\beta_{ij}$.}

\figcaption[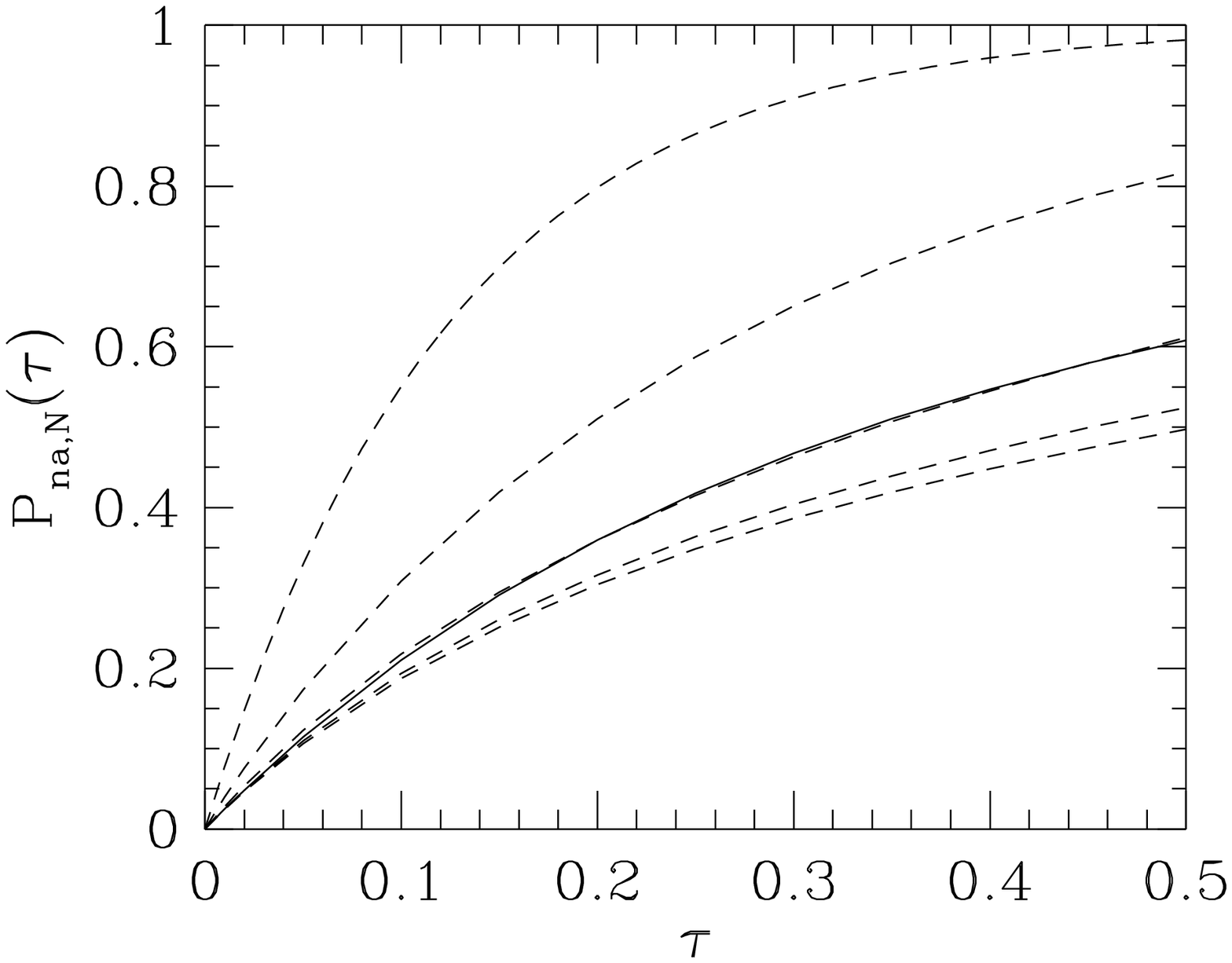]{The fraction of point masses whose caustics are not
isolated astroids as a function of the optical depth $\tau$.
The dashed lines from top to bottom are $P'_{{\rm na},N}$ (eq.[\ref{eq36}]) for
$N = 1$, $2$, $4$, $8$, and $\infty$, and the solid line is
$P_{{\rm na},\infty}$ (eq.[\ref{eq37}]).
A lens distribution with constant comoving mass density is assumed
(see \S~4.2).}

\clearpage

\plotone{fig1.ps}

\clearpage

\plotone{fig2.ps}

\clearpage

\plotone{fig3ab.ps}

\clearpage

\plotone{fig3c.ps}

\clearpage

\plotone{fig4.ps}

\clearpage

\plotone{fig5.ps}

\clearpage

\plotone{fig6ab.ps}

\clearpage

\plotone{fig6cd.ps}

\clearpage

\plotone{fig6ef.ps}

\clearpage

\plotone{fig7.ps}

\clearpage

\plotone{fig8.ps}

\clearpage

\plotone{fig9.ps}

\clearpage

\plotone{fig10ab.ps}

\clearpage

\plotone{fig10c.ps}

\clearpage

\plotone{fig11.ps}

\clearpage

\plotone{fig12.ps}

\clearpage

\plotone{fig13.ps}

\clearpage

\plotone{fig14.ps}

\clearpage

\plotone{fig15.ps}

\end{document}